\documentclass[reprint,superscriptaddress,showpacs,preprintnumbers,amsmath,amssymb,aps,longbibliography,prb,floatfix]{revtex4-1}

\usepackage[english]{babel}
\usepackage{multirow}
\usepackage[thinspace]{SIunits}
\usepackage{xspace}
\usepackage{graphicx}
\usepackage{dcolumn}
\usepackage{bm}
\usepackage{color}
\usepackage{hyperref}
\hypersetup{unicode=false,pdftoolbar=true,pdfmenubar=true,pdffitwindow=false,pdfstartview={FitH},pdftitle={My title},pdfauthor={Author},pdfsubject={Subject},pdfcreator={Creator},pdfproducer={Producer},pdfkeywords={keyword1} {key2} {key3},pdfnewwindow=true,colorlinks=true,linkcolor=blue,citecolor=blue,filecolor=blue,urlcolor=blue,plainpages=false}

\graphicspath{{./}}

\newcommand{\TSDW}{\ensuremath{T_\mathrm{SDW}}\xspace}
\newcommand{\Ts}{\ensuremath{T_\mathrm{s}}\xspace}
\newcommand{\BFCA}{\mbox{Ba(Fe$_{\mathrm{1-x}}$Co$_{\mathrm{x}}$)$_{2}$As$_2$}\xspace}
\newcommand{\BFA}{\mbox{BaFe$_{2}$As$_2$}\xspace}
\newcommand{\Alg}{\ensuremath{A_{1g}}\xspace}
\newcommand{\Ag}{\ensuremath{A_{g}}\xspace}

\newcommand{\Blg}{\ensuremath{B_{1g}}\xspace}
\newcommand{\BZg}{\ensuremath{B_{2g}}\xspace}
\newcommand{\BEg}{\ensuremath{B_{3g}}\xspace}
\newcommand{\Eg}{\ensuremath{E_{g}}\xspace}

\newcommand{\vei}{\mbox{{\bf e}$_\mathrm{I}$}\xspace}
\newcommand{\ves}{\mbox{{\bf e}$_\mathrm{S}$}\xspace}
\newcommand{\wi}{\ensuremath{\omega_\mathrm{I}}\xspace}
\newcommand{\ws}{\ensuremath{\omega_\mathrm{S}}\xspace}
\newcommand{\wn}{\ensuremath{\mathrm{cm}^{-1}}\xspace}
\begin{document}

\title{Interplay of lattice, electronic and spin degrees of freedom in detwinned \texorpdfstring{\BFA}{BaFe2As2}: \\ a Raman scattering study}
\date{\today}
\author{A. Baum}
\affiliation{Walther Meissner Institut, Bayerische Akademie der Wissenschaften, 85748
Garching, Germany}
\affiliation{Fakult\"at f\"ur Physik E23, Technische Universit\"at M\"unchen, 85748
Garching, Germany}
\author{Ying Li}
\affiliation{Institut f\"ur Theoretische Physik, Goethe-Universit\"at Frankfurt,
Max-von-Laue-Stra{\ss }e 1, 60438 Frankfurt am Main, Germany}
\author{M.~Tomi\'c}
\affiliation{Institut f\"ur Theoretische Physik, Goethe-Universit\"at Frankfurt,
Max-von-Laue-Stra{\ss }e 1, 60438 Frankfurt am Main, Germany}
\author{N. Lazarevi\'c}
\affiliation{Center for Solid State Physics and New Materials, Institute of Physics
Belgrade, University of Belgrade, Pregrevica 118, 11080 Belgrade, Serbia}
\author{D. Jost}
\affiliation{Walther Meissner Institut, Bayerische Akademie der Wissenschaften, 85748
Garching, Germany}
\affiliation{Fakult\"at f\"ur Physik E23, Technische Universit\"at M\"unchen, 85748
Garching, Germany}
\author{F. L\"offler}
\affiliation{Walther Meissner Institut, Bayerische Akademie der Wissenschaften, 85748
Garching, Germany}
\affiliation{Fakult\"at f\"ur Physik E23, Technische Universit\"at M\"unchen, 85748
Garching, Germany}
\author{B.~Muschler}
\altaffiliation{Present address: Zoller \& Fr\"ohlich GmbH, Simoniusstrasse 22, 88239 Wangen
im Allg\"au,Germany}
\affiliation{Walther Meissner Institut, Bayerische Akademie der Wissenschaften, 85748
Garching, Germany}
\affiliation{Fakult\"at f\"ur Physik E23, Technische Universit\"at M\"unchen, 85748
Garching, Germany}
\author{T.~B\"ohm}
\altaffiliation{Present address: TNG Technology Consulting GmbH, Beta-Stra\ss{}e, 85774 Unterf\"{o}hring, Germany}
\affiliation{Walther Meissner Institut, Bayerische Akademie der Wissenschaften, 85748
Garching, Germany}
\affiliation{Fakult\"at f\"ur Physik E23, Technische Universit\"at M\"unchen, 85748
Garching, Germany}
\author{J.-H. Chu}
\affiliation{Stanford Institute for Materials and Energy Sciences, SLAC National
Accelerator Laboratory, 2575 Sand Hill Road, Menlo Park, CA 94025, USA}
\affiliation{Geballe Laboratory for Advanced Materials \& Dept. of Applied Physics,
Stanford University, CA 94305, USA}
\affiliation{Department of Physics, University of Washington, Seattle WA 98195, USA}
\author{I.~R. Fisher}
\affiliation{Stanford Institute for Materials and Energy Sciences, SLAC National
Accelerator Laboratory, 2575 Sand Hill Road, Menlo Park, CA 94025, USA}
\affiliation{Geballe Laboratory for Advanced Materials \& Dept. of Applied Physics,
Stanford University, CA 94305, USA}
\author{R. Valent\'i}
\affiliation{Institut f\"ur Theoretische Physik, Goethe-Universit\"at Frankfurt,
Max-von-Laue-Stra{\ss }e 1, 60438 Frankfurt am Main, Germany}
\author{I.\,I.~Mazin}
\affiliation{Code 6393, Naval Research Laboratory, Washington, DC 20375, USA}
\author{R. Hackl}
\email{hackl@wmi.badw.de}
\affiliation{Walther Meissner Institut, Bayerische Akademie der Wissenschaften, 85748
Garching, Germany}

\begin{abstract}
We report results of Raman scattering experiments on twin-free \BFA with the main focus placed on understanding the influence of electronic and spin degrees of freedom on the lattice dynamics. In particular, we scrutinize the \Eg modes and the As \Alg mode. Each of the two \Eg phonons in the tetragonal phase is observed to split into a \BZg and a \BEg mode upon entering the orthorhombic stripe-magnetic phase. The splitting amounts to approximately 10\,cm$^{-1}$ and less than 5\,cm$^{-1}$ for the low- and the high-energy \Eg mode, respectively. The detailed study of the fully symmetric As mode using parallel incident and outgoing photon polarizations along either the antiferromagnetic or the ferromagnetic Fe-Fe direction reveals an anisotropic variation of the spectral weight with the energy of the exciting laser indicating a polarization-dependent resonance effect.
Along with the experiments we present results from density functional theory calculations of the phonon eigenvectors, the dielectric function, and the Raman tensor elements. The comparison of theory and experiment indicates that (i) orbital-selective electronic correlations are crucial to understand the lattice dynamics and (ii) all phonon anomalies originate predominantly from the magnetic ordering and the corresponding reconstruction of the electronic bands at all energies.
\end{abstract}

\maketitle


\section{Introduction}

One of the most debated issues in Fe-based superconductors is the interplay of spin, orbital and lattice degrees of freedom at the onset of magnetism, nematicity and superconductivity. \cite{Sefat:2011,Wang:2015,Gallais:2016a,YiM:2017,Bohmer:2018} Actually, phonons may play a decisive role for probing subtle changes of the electronic and magnetic properties. For instance, soon after the discovery of Fe-based superconductors the magnetic moment was predicted to couple to the As position. \cite{Yildirim:2009a} Zbiri \textit{et al.} found a modulation of the electronic density of states at the Fermi energy $E_\mathrm{F}$ by the two \Eg and the \Alg modes. \cite{Zbiri:2009i} Various anomalies were observed experimentally using neutron, Raman and optical spectroscopy, \cite{Chauviere:2009,Chauviere:2011,Rahlenbeck:2009,Kumar:2010,Mittal:2009,Gnezdilov:2013,Gnezdilov:2011,Akrap:2009} but are not fully understood yet.

\begin{figure}[tbh]
\centering
\includegraphics[width=8.5cm]{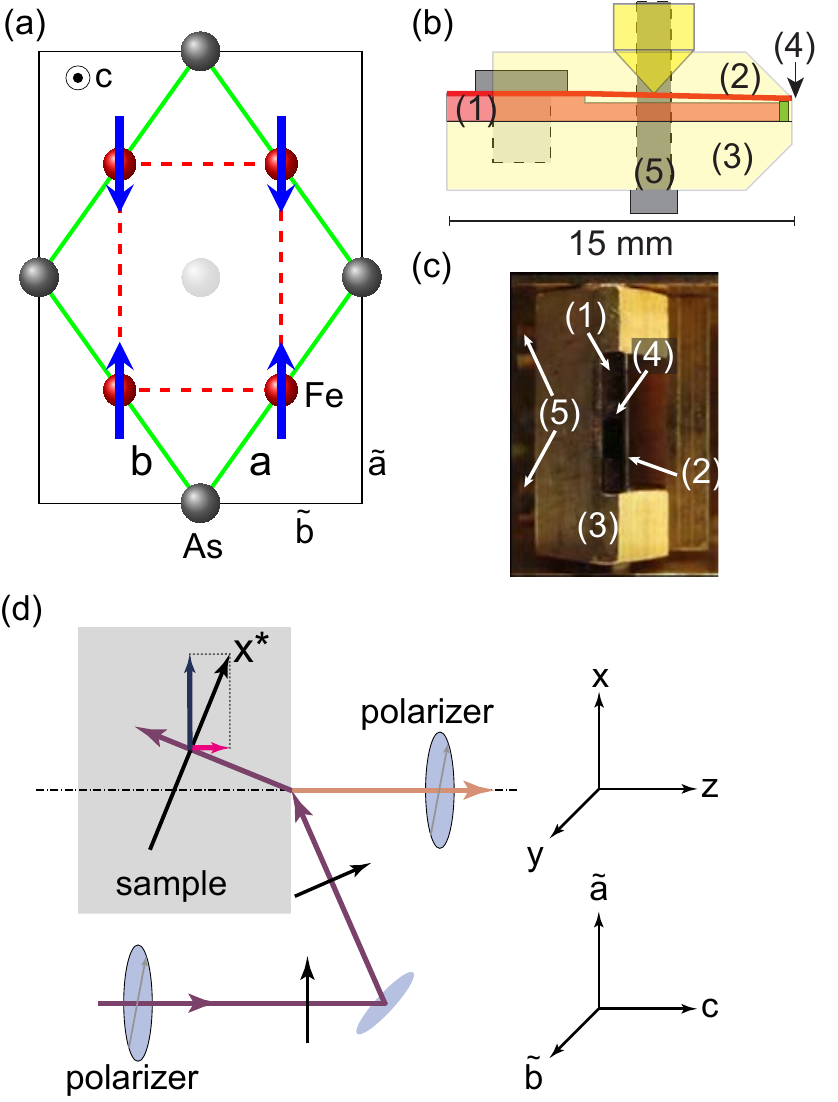}  
\caption{(Color online) FeAs layer of \BFA and detwinning clamp. (a) The As-atoms (grey) in the center and at the edges are below and, respectively, above the Fe plane (red). For this reason, the 2\,Fe unit cell with the axes $a$ and $b$ (green) is determined by the As atoms. In the orthorhombic phase the Fe-Fe distances become inequivalent with the distortion strongly exaggerated here. The magnetic unit cell is twice as large as the 2\,Fe unit cell and has the axes $\tilde{a}$ and $\tilde{b}$.
(b) Schematic sketch and (c) photograph of the detwinning clamp. The sample (4) is glued on the copper plate (1) which is in good thermal contact with the sample holder (3). Upon tightening the screws (5) the force exerted by the copper-beryllium cantilever (2) can be adjusted. (d) Schematic representation of the geometry of our Raman scattering experiment. All incoming light polarizations \vei which are not parallel to $y$ have finite projections on the $c$ axis (red arrow).}
\label{fig:orth_phase}
\end{figure}

One particular effect is the observation of substantial Raman scattering intensity of the As phonon below the magneto-structural transition in crossed polarizations with the electric fields oriented along the axes of the pseudo-tetragonal 2\,Fe unit cell \cite{Chauviere:2011} [For the definition of the axes see Fig.~\ref{fig:orth_phase}\,(a)]. Garc\'ia-Mart\'{\i}nez \textit{et al.} argued that magnetism sufficiently modifies the low-energy electronic structure to explain this anomalous intensity.\cite{Garcia:2013} Recent experiments seem to support this view \cite{Wu:2017dec} upon comparing spectra obtained with parallel and crossed polarizations in twinned samples with the incident field oriented along the $a$ and the scattered field either along the $a$ or $b$ axis, respectively. Yet, to which extent the phonons are affected by correlations and magnetic-ordering-induced changes in the electronic structure at energies in the range of the photon energies is still unclear.

In this work we address this issue both experimentally and theoretically and investigate how magnetism and the combination of moderately correlated Fe $d$ states and uncorrelated As $p$ states affect such complex spectroscopic properties as, for instance, resonant Raman scattering. In particular, we try to clarify whether the observed anomalous intensity of the As mode is a low- or a high-energy phenomenon and aim at identifying the driving force behind the ordering instabilities.

In our study we find that very good agreement between experimental observations and  density functional theory (DFT) calculations can be achieved in both the para\-magnetic and the antiferromagnetic state of \BFA if two physically motivated modifications are being made to the standard DFT electronic bands. On the one hand,  we need to account for the fact that the high-temperature tetragonal phase is paramagnetically disordered, and cannot be simulated by calculations with suppressed local magnetism.\cite{Mazin:2008a} Besides, it appears necessary not only to introduce an antiferromagnetic order in the calculations, but also to account for strong correlations. The latter is achieved by separating the energy bands into two regions, a high-energy region with predominantly As states and a low-energy region with  predominantly Fe states. The Fe states are then appropriately renormalized. With these two assumptions we can reproduce (i) the positions of the Raman active phonons and their splitting and evolution in the (mechanically detwinned) orthorhombic antiferromagnetic state and (ii) Raman intensities, including the $\tilde{a}-\tilde{b}$ anisotropy as well as the complex resonant evolution with the laser light frequency. This agreement gives an experimental justification to the proposed computational procedure and convincingly substantiates the physical concepts it was derived from, namely the pivotal role of local moments in the lattice dynamics of Fe-based superconductors, and the importance of band renormalizations for $d$-electrons.


\section{Methods}

\subsection{Samples}

\label{sec:samples}

The \BFA crystal was prepared using a self-flux technique. Details of the crystal growth and characterization are described elsewhere.\cite{Chu:2009} \BFA is a parent compound of double-layer iron-based superconductors and orders in a stripe-like spin-density-wave (SDW) below $\TSDW \approx \mathrm{135\,K}$. Superconductivity can be obtained by substituting any of the ions or by pressure.\cite{Kimber:2009i} In \BFCA ($0<x\lesssim 0.06$) the SDW is preceded by a structural phase transition from a tetragonal ($I4/mmm$) to an orthorhombic ($Fmmm$) lattice at $\Ts > \TSDW$.\cite{Chu:2009} It remains a matter of debate as to whether or not \TSDW and \Ts coincide in \BFA.\cite{Chu:2009,Kim:2011b}

Fig.~\ref{fig:orth_phase}(a) shows the relation of the various axes. The axes of the tetragonal crystal ($T > T_{\mathrm{s}}$, green lines) are denoted $a$ and $b$ with $a = b$. The axes of the magnetically ordered structure (4\,Fe per unit cell, black lines), $\tilde{a}$ and $\tilde{b}$, differ by approximately 0.7\% below \TSDW\cite{Rotter:2008} and the Fe-Fe distance along the $\tilde{b}$ axis becomes shorter than along the $\tilde{a}$ axis as sketched in Figure~\ref{fig:orth_phase}(a). As a result, the angle between $a$ and $b$ differs from 90$^{\circ}$ by approximately 0.4$^{\circ}$.

Below \TSDW the spins order ferromagnetically along $\tilde{b}$ and antiferromagnetically along $\tilde{a}$. Due to the small difference between $\tilde{a}$ and $\tilde{b}$ the crystals are twinned below \Ts, and the orthogonal $\tilde{a}$ and $\tilde{b}$ axes change roles at twin boundaries running along the directions of the tetragonal $a$ and $b$ axes. The orthorhombic distortion makes the proper definition of the axes important as has been shown for twin-free crystals by longitudinal and optical transport as well as by ARPES.\cite{Chu:2010,Ying:2011,Dusza:2011,Dusza:2012,Nakajima:2011,Yi:2011} In order to obtain a single-domain orthorhombic crystal we constructed a sample holder for applying uniaxial pressure parallel to the Fe-Fe direction.

\subsection{Detwinning clamp}

\label{sec:clamp}

The detwinning clamp is similar to that used by Chu \textit{et al.}\cite{Chu:2010} Fig. \ref{fig:orth_phase}(b) and (c) show, respectively, a schematic drawing and a photograph of the clamp. The sample is attached to a thermally sinked copper block (1) with GE varnish, which remains sufficiently elastic at low temperatures and maintains good thermal contact between the holder (3) and the sample (4). The stress is applied using a copper-beryllium cantilever (2) which presses the sample against the body of the clamp. Upon tightening the screws (5) the force on the sample can be adjusted. In our experiment, the pressure is applied along the Fe-Fe bonds. The $c$ axis of the sample is perpendicular to the force and parallel to the optical axis. The uniaxial pressure can be estimated from the rate of change of the tetragonal-to-orthorhombic phase transition at $T_{\mathrm{s}}$. Using the experimentally derived rate of \SIunits{1}{\usk\kelvin} per \SIunits{7}{\usk\mega\pascal} \cite{Liang:2011,Blomberg:2012} we find approximately \SIunits{35}{\usk\mega\pascal} for our experiment to be sufficient to detwin the sample.

\subsection{Light scattering}

\label{sec:ls}

The experiment was performed with a standard light scattering setup. We used two ion lasers (Ar$^{+}$ Coherent Innova 304C and Kr$^{+}$ Coherent Innova 400) and two diode pumped solid state lasers (Coherent Genesis MX SLM, Laser Quantum Ignis) providing a total of 14 lines ranging from \SIunits{407}{\usk\nano\meter} to \SIunits{676}{\usk\nano\meter}, corresponding to incident energies $\hbar\wi$ between 3.1 and 1.8\,eV. Due to this wide range the raw data have to be corrected.
The quantity of interest is the response function $R\chi^{\prime\prime}(\Omega)$ where $\Omega=\wi-\ws$ is the Raman shift, \ws is the energy of the scattered photons and $R$ is an experimental constant. Details of the calibration are described in Appendix~\ref{asec:dataevaluation}.

Application of the Raman selection rules requires well-defined polarizations for the exciting and scattered photons. The polarizations are given in Porto notation with the first and the second symbol indicating the directions of the incoming and scattered photons' electric fields \vei and \ves, respectively. We use $xyz$ for the laboratory system [see Fig. \ref{fig:orth_phase}(d)]. The $xz$ plane is vertical and defines the plane of incidence, $yz$ is horizontal, $xy$ is the sample surface, and the $z$ axis is parallel to the optical axis and to the crystallographic $c$ axis. For the sample orientation used here (see Fig.~\ref{fig:orth_phase}) the Fe-Fe bonds are parallel to $x$ and $y$, specifically $\tilde{a} = (1,0,0) \parallel x$ and $\tilde{b} = (0,1,0) \parallel y$. Since the orthorhombicity below \Ts is small the angle between $a$ and $\tilde{a}$ deviates only by 0.2$^{\circ}$ from 45$^{\circ}$. It is therefore an excellent approximation to use $a \parallel x^{\prime}=1/\sqrt{2}(x+y) \equiv 1/\sqrt{2}(1,1,0)$ and $b \parallel y^{\prime}=1/\sqrt{2}(y-x) \equiv 1/\sqrt{2}(1,\bar{1},0)$.

As the angle of incidence of the exciting photons is as large as 66$^{\circ}$ in our setup [see Fig. \ref{fig:orth_phase}(d)] the orientations of \vei parallel and perpendicular to the $xz$
plane are inequivalent. In particular, \vei has a projection on the $c$ axis for $\vei \parallel xz$. This effect was used before\cite{Chauviere:2009} and allows one to project out the \Eg phonons in the $x^{\ast}x$ and $x^{\ast}y$ configurations, where $x^{\ast} \parallel (x+\alpha z)$ inside the crystal [see Fig. \ref{fig:orth_phase}(d)]. For \BFA the index of refraction is $n^{\prime} = 2.2+2.1i$ at 514\,nm resulting in $\alpha \approx 0.4$ for an angle of incidence of 66$^{\circ}$. The corresponding intensity contribution is then 0.16. As a consequence, $x^{\ast}x$ and $yy$ are inequivalent whereas $\vei = x^{\prime\ast} \parallel (x^{\prime} + \alpha\,z/\sqrt{2})$ and $\vei = y^{\prime\ast} \parallel (y^{\prime} + \alpha\,z/\sqrt{2})$ are equivalent for having the same projection on the $c$ direction. Upon comparing $x^{\ast}y$ and $yx$ the leakage of the $c$-axis polarized contributions to the electronic continuum can be tested. In the case here, they are below the experimental sensitivity. The effect of the finite angle of acceptance of the collection optics ($\pm 15^{\circ}$ corresponding to a solid angle $\tilde{\Omega}$ of 0.21\,sr) on the projections of the scattered photons can be neglected.

\begin{figure}[tbp]
\centering
\includegraphics[width=8.5cm]{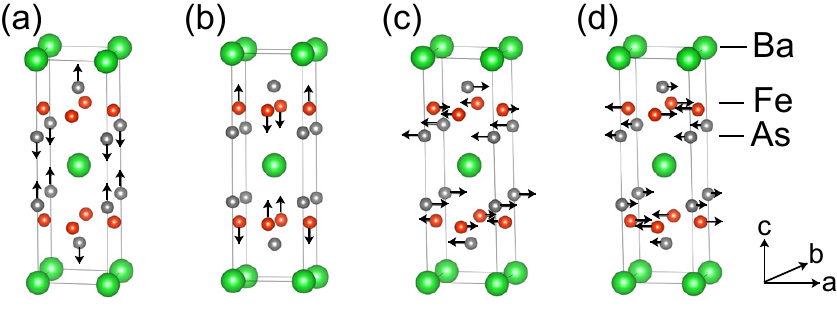}  
\caption{(Color online) Raman-active phonons in \BFA with the symmetry assignments in the tetragonal crystallographic unit cell $abc$.}
\label{fig:mode_tetra}
\end{figure}

\subsection{Theoretical Calculations}
\label{sec:theory} 

The phonon eigenvectors $Q^{(\nu)}$ (displacement patterns of the vibrating atoms in branch $\nu$) and the energies of all Raman-active phonons of \BFA in the tetragonal ($I4/mmm$) and the orthorhombic ($Fmmm$) phases were obtained from \textit{ab initio} DFT calculations within the Perdew-Burke-Ernzerhof parameterization \cite{Perdew:1996} of the generalized gradient approximation. The phonon frequencies were calculated by diagonalizing the dynamical matrices using the \textit{phonopy} package. \cite{Togo:2008,Togo:2015} The dynamical matrices were constructed from force constants determined via the finite displacement method in 2 $\times$ 2 $\times$ 1 supercells. \cite{Parlinski:1997} As a basis for the calculations we used the projector
augmented wave approximation, \cite{Bloechl:1994} as implemented in the Vienna package (VASP). \cite{Kresse:1993,Kresse:1996,Kresse:1996b} The Brillouin zone for one unit cell was sampled with a $10 \times 10 \times 10$ \textbf{k} point mesh, and the plane wave cutoff was set at 520\,eV. For the tetragonal phase, we used a N\'{e}el-type magnetic order to relax the structure and to obtain the experimental lattice parameters. \footnote{The local correlations in the tetragonal phase are of the stripe type; however, we had to use a pattern that does not break the symmetry, and it is known \cite{Mazin:2008a} that the difference in the calculated elastic properties calculated within different magnetic orders is much smaller than between magnetic and nonmagnetic calculations} For the orthorhombic phase, we used the stripe-like magnetic order shown in Fig.~\ref{fig:orth_phase}(a).

In addition, we studied the resonant phonon-photon interaction by exploring the dielectric tensor $\hat{\varepsilon}$. The latter was determined using the Optics code package \cite{Ambrosch-Draxl:2006} implemented in WIEN2k (Ref.~\onlinecite{Blaha:2001}) with the full-potential linearized augmented plane-wave (LAPW) basis. The Perdew-Burke-Ernzerhof generalized gradient approximation \cite{Perdew:1996} was employed as the exchange correlation functional and the basis-size controlling parameter RK$_{\mathrm{max}}$ was set to 8.5. A mesh of 400 \textbf{k} points in the first Brillouin zone for the self-consistency cycle was used. The density of states (DOS) and dielectric tensors were computed using a $10\times10\times10$ \textbf{k} mesh. For the dielectric tensor a Lorentzian broadening of 0.1\,eV was introduced.

The (generally complex) Raman tensor ${\alpha}_{jk}^{(\nu)}(\wi) = \alpha^{(\nu)\prime}_{jk}(\wi) + i\alpha^{(\nu)\prime\prime}_{jk}(\wi)$ is determined by the derivative of the dielectric tensor elements ${\varepsilon}_{jk}(\wi)=\varepsilon^\prime_{jk}(\wi) + i\varepsilon^{\prime\prime}_{jk}(\wi)$ with respect to the normal coordinate of the respective phonon, $Q^{(\nu)}$.
Since we are interested only in the resonance behavior of the As phonon, we are only concerned with the derivative with respect to $Q^{\mathrm{(As)}}$, 
\begin{equation}  \label{eq:de_dQ}
{\alpha}_{ll}^{\mathrm{(As)}}(\wi) = \frac{\partial\varepsilon^\prime_{ll}(\wi)}{\partial Q^{\mathrm{(As)}}} + i\frac{\partial\varepsilon^{\prime\prime}_{ll}(\wi)}{\partial Q^{\mathrm{(As)}}}.
\end{equation}


\section{Results and Discussion}
\label{sec:results}

\begin{table}[tbp]
\caption{Raman-active phonons in \BFA. The experimental and theoretically determined energies are given in cm$^{-1}$. In addition, the symmetry correlations between the tetragonal ($I4/mmm$) and orthorhombic ($Fmmm$) structures are shown.}
\label{table:phonon}
\renewcommand{\arraystretch}{1.6}  
\begin{ruledtabular}
		\begin{tabular}{lccclcc}
		\multicolumn{3}{c}{$I4/mmm$} & & \multicolumn{3}{c}{$Fmmm$} \\
		& Exp. (140\,K) & Theory & & & Exp. (60\,K) & Theory\\
		\hline
		\Alg &180 &168        & $\xrightarrow{\makebox[5mm]{}}$   &   \Ag &180  &172   \\
		\Blg &215 &218        & $\xrightarrow{\makebox[5mm]{}}$   &   \Blg &215 &221   \\
		\multirow{2}{*}{$\Eg^{(1)}$}      & \multirow{2}{*}{130} & \multirow{2}{*}{140} &\multirow{ 2}{*}{{\begin{tabular}{c}\rotatebox[origin=l]{10}{$\xrightarrow{\hspace*{5mm}}$} \\[-5mm] \rotatebox[origin=l]{-10}{$\xrightarrow{\hspace*{5mm}}$} \end{tabular}           }}&   $\BZg^{(1)}$ &125 &110      \\
																 &&&&                                                                $\BEg^{(1)}$ &135 &133      \\
	\multirow{2}{*}{$\Eg^{(2)}$} &\multirow{2}{*}{268}  &\multirow{2}{*}{290}         &\multirow{ 2}{*}{{\begin{tabular}{c}\rotatebox[origin=l]{10}{$\xrightarrow{\hspace*{5mm}}$} \\[-5mm] \rotatebox[origin=l]{-10}{$\xrightarrow{\hspace*{5mm}}$} \end{tabular}           }}&   $\BZg^{(2)}$ &270 &272      \\
																 &&&&                                                                $\BEg^{(2)}$& 273 &287       \\
		\end{tabular}
  \end{ruledtabular}
\end{table}

\subsection{Lattice dynamics}
\label{sec:phononcalc}

The energies and symmetries as obtained from lattice dynamical calculations for tetragonal and orthorhombic \BFA are compiled in Table~\ref{table:phonon}. The four modes in tetragonal $I4/mmm$ symmetry obey \Alg + \Blg + 2\,\Eg selection rules.
The eigenvectors are depicted in Fig.~\ref{fig:mode_tetra}. 
In the orthorhombic $Fmmm$ phase, the two \Eg modes are expected to split into \BZg and \BEg modes. Thus, there are six non-degenerate modes in the orthorhombic phase, \Ag + \Blg + 2\,\BZg + 2\,\BEg. Table~\ref{table:phonon} shows the symmetry relations between the tetragonal and orthorhombic phonons.

\begin{figure}[tbp]
\centering-
\includegraphics[width=8.5cm]{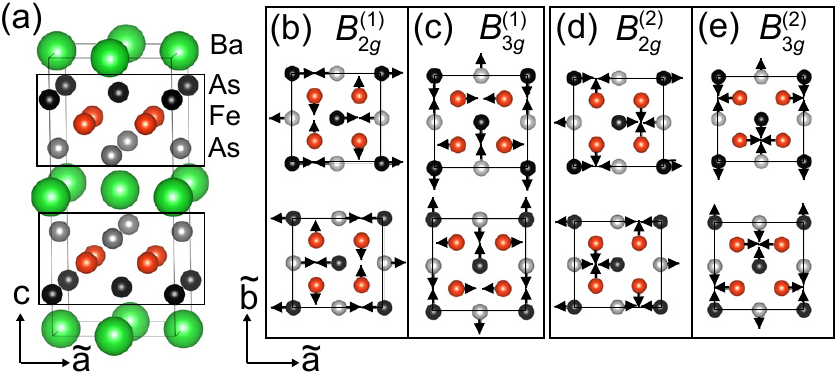} 
\caption{(Color online) \BZg and \BEg phonon modes in \BFA with the symmetry assignments in the orthorhombic crystallographic unit cell $\tilde{a}\tilde{b}c$. }
\label{fig:mode_orthor}
\end{figure}

Since the \Ag and \Blg eigenvectors remain unchanged upon entering the orthorhombic phase, only those of the \BZg and \BEg phonons are shown in Fig.~\ref{fig:mode_orthor}. For the \BZg and \BEg phonons the As and Fe atoms move perpendicular to the $c$ axis and perpendicular to each other. The calculated phonon vibrations agree with previous results for \BFA, \cite{Zbiri:2009i} however our energies differ slightly from those reported by Zbiri \textit{et al.} \cite{Zbiri:2009i} In particular, we find a splitting between the $\BZg^{(1)}$ and $\BEg^{(1)}$ phonons.

\subsection{\texorpdfstring{\Eg}{Eg} phonons}
\label{sec:Egphonon}

\begin{figure}[tbp]
\centering
\includegraphics[width=8.5cm]{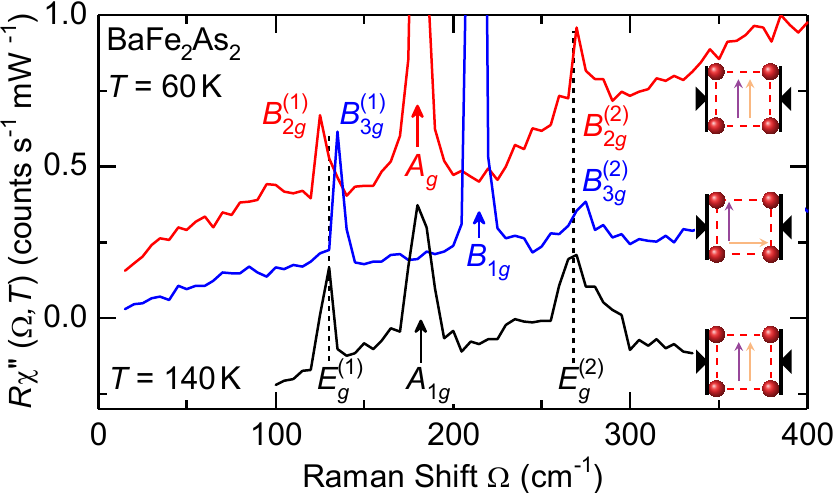}  
\caption{(Color online) Phonons in detwinned \BFA. The spectra at 60\,K (red and blue) are displayed with the experimental intensity. The spectrum at 140\,K (black) is downshifted by 1.4\,counts\,s$^{-1}$\,mW$^{-1}$ for clarity. Each of the two tetragonal \Eg phonons (vertical dashed lines) splits into two lines below \Ts. The $\BZg^{(1)}$ and $\BEg^{(1)}$ lines appear at distinct positions for polarizations of the scattered light parallel (blue) and perpendicular (red) to the applied pressure as indicated in the insets. The $\BZg^{(2)}$ and $\BEg^{(3)}$ phonons are shifted only slightly upwards with respect to the $\Eg^{(2)}$ mode. Violet and orange arrows indicate the polarizations of the incident and scattered photons, respectively. The black triangles indicate the direction of the applied pressure. The shorter $\tilde{b}$ axis is parallel to the stress.}
\label{fig:Eg_Phonon_Split}
\end{figure}

Table~\ref{table:phonon} displays the experimental phonon energies as measured above and below the magneto-structural transition along with the theoretical values. The $\Eg^{(1)}$ phonon found at 130\,cm$^{-1}$ above \Ts splits into two well separated lines as predicted (Table~\ref{table:phonon}) and shown in Fig.~\ref{fig:Eg_Phonon_Split}. The splitting of the $\Eg^{(2)}$ mode at 268\,cm$^{-1}$ is small, and the $\BZg^{(2)}$ and $\BEg^{(2)}$ modes are shifted to higher energies by 2\,\wn, and 5\,cm$^{-1}$, respectively.

The theoretical and experimental phonon energies are in agreement to within 14\% for both crystal symmetries. The splitting between the \BZg and \BEg modes is overestimated in the calculations.

Previous experiments were performed on twinned crystals, \cite{Chauviere:2009,RenX:2015,ZhangWL:2016} and the \BZg and \BEg modes were observed next to each other in a single spectrum. An equivalent result can be obtained in de-twinned samples by using $x^{\prime \ast}y^{\prime}$, $x^{\prime \ast}x^{\prime}$ or $RR$ polarizations where the $x$ and $y$ axes are simultaneously projected (along with the $z$ axis). In neither case the symmetry of the \BZg and \BEg phonons can be pinned down. Only in a de-twinned sample where the $xz$ and $yz$ configurations are projected separately the \BZg and \BEg modes can be accessed independently.

Uniaxial pressure along the Fe-Fe direction, as shown by the black arrows in the insets of Fig.~\ref{fig:Eg_Phonon_Split}, determines the orientation of the shorter $\tilde{b}$ axis. This configuration enables us to observe the $\BZg^{(1)}$ mode at 125\,cm$^{-1}$ and the $\BEg^{(1)}$ mode at 135\,cm$^{-1}$ in $x^{\ast}x$ and, respectively, $x^{\ast}y$ polarization configurations thus augmenting earlier work. With the shorter axis determined by the direction of the stress (insets of Fig.~\ref{fig:Eg_Phonon_Split}) the assignment of the \BZg and \BEg modes is unambiguous. Since the $x^{\ast}x$ spectrum (red) comprises $\tilde{a}\tilde{a}$ and $c\tilde{a}$ polarizations both the \Ag and the \BZg phonons appear. The $x^{\ast}y$ spectrum (blue) includes the \Blg ($\tilde{a}\tilde{b}$) and \BEg ($c\tilde{b}$) symmetries.

The calculated splitting between the \BZg and \BEg modes is smaller for the $\Eg^{(2)}$ than for the $\Eg^{(1)}$ mode, qualitatively agreeing with the experiment. However, in the calculations this difference is entirely due to the different reduced masses for these modes since the $\Eg^{(1)}$ and $\Eg^{(2)}$ phonon are dominated by As and Fe motions, respectively. In the experiment the splitting for the $\Eg^{(2)}$ mode is close to the spectral resolution, indicating an additional reduction of the splitting below that obtained in the calculation. The source of this additional reduction is unclear at the moment.

\subsection{As phonon intensity}
\label{sec:Agphonon}

\begin{figure}[tbp]
\centering
\includegraphics[width=8.5cm]{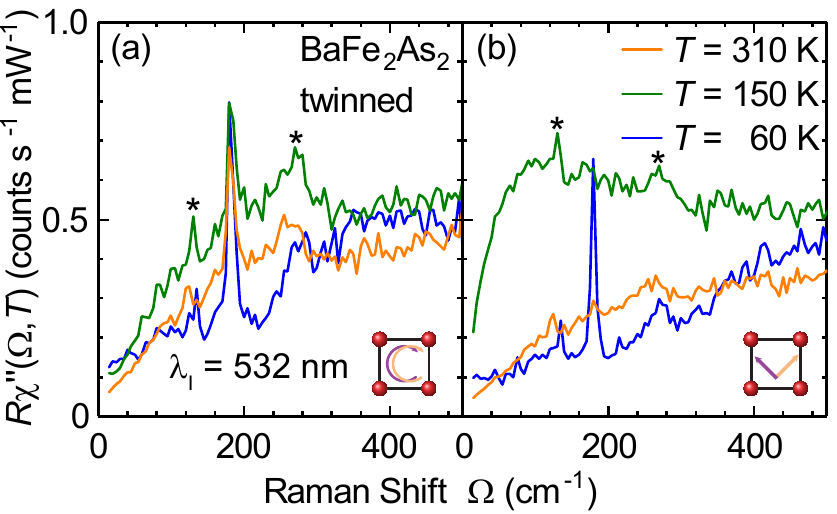}  
\caption{(Color online) Raman spectra of twinned \BFA at temperatures as indicated. (a) In parallel $RR$ polarization configuration (see inset) the As phonon appears at all temperatures. (b) For crossed light polarizations ($ab$) the As phonon is present only below the magneto-structural transition at $\Ts = 135$\,K as reported before. \cite{Chauviere:2009} Asterisks mark the \Eg modes discussed in Sec.~\ref{sec:Egphonon}.}
\label{fig:Ag_Phonon_in_B1g}
\end{figure}

Fig.~\ref{fig:Ag_Phonon_in_B1g} shows low-energy spectra of twinned \BFA for (a) $RR$ and (b) $ab$ polarization configurations at 310 (orange), 150 (green), and 60\,K (blue). The As phonon at 180\,\wn is the strongest line in the $RR$ spectra at all temperatures as expected and gains intensity upon cooling. In $ab$ polarizations there is no contribution from the As mode above \Ts. Below \Ts (blue spectrum) the As phonon assumes a similar intensity as in the $RR$ polarization as reported earlier. \cite{Chauviere:2009,Wu:2017dec} Due to a finite projection of the incident light polarizations onto the $c$ axis [see Fig.~\ref{fig:orth_phase}(d)] in both $RR$ and $ab$ configurations the \Eg phonons appear in all spectra (asterisks). The electronic background has been extensively discussed in previous works~\cite{Choi:2008,Chauviere:2010,Sugai:2012,Kretzschmar:2016,Thorsmolle:2016} and is not a subject of the study here.

In order to understand the appearance of the As line in the crossed $ab$ polarizations it is sufficient to consider the in-plane components of the \Ag Raman tensor, 
\begin{equation}  
\label{eq:tensor-Ag}
\hat{\alpha}^{(Ag)} = \left( 
\begin{array}{cc}
{\alpha}_{11} & 0 \\ 
0 & {\alpha}_{22}%
\end{array}
\right).
\end{equation}

The response of this phonon for the polarization configuration (\vei,\ves) is given by $\chi^{\prime\prime \mathrm{(As)}}_{IS} \propto \left| \mathbf{e}_{\mathrm{S}%
}^{\ast} \cdot \hat{\alpha}^{(Ag)} \cdot \vei \right|^2$ (where $^\ast$ means conjugate transposed). In the tetragonal (\Alg) case the two elements are equal, ${\alpha}_{11} = {\alpha}%
_{22}$, and the phonon appears only for $\ves \! \parallel\! \vei$. In the orthorhombic
phase the tensor elements are different, and one can expect the phonon to appear for $\ves \perp \vei$ since the intensity then depends on the difference between $\alpha_{11}$ and $\alpha_{22}$. In detwinned samples $\alpha_{11}$ and $\alpha_{22}$ can be accessed independently by using parallel polarizations for the incident and scattered light oriented along either the  $\tilde{a}$ or the $\tilde{b}$ axis. In addition, putative imaginary parts of $\alpha_{ii}$ may be detected by analyzing more than two polarization combinations as discussed in Appendix~\ref{asec:complex_tensor}. Spectra for $\tilde{a}\tilde{a}$ and $\tilde{b}\tilde{b}$ configurations are shown in Figure~\ref{Afig:Ag_all} of Appendix~\ref{asec:rawdata}.

\begin{figure}[tbp]
\centering
\includegraphics[width=8.5cm]{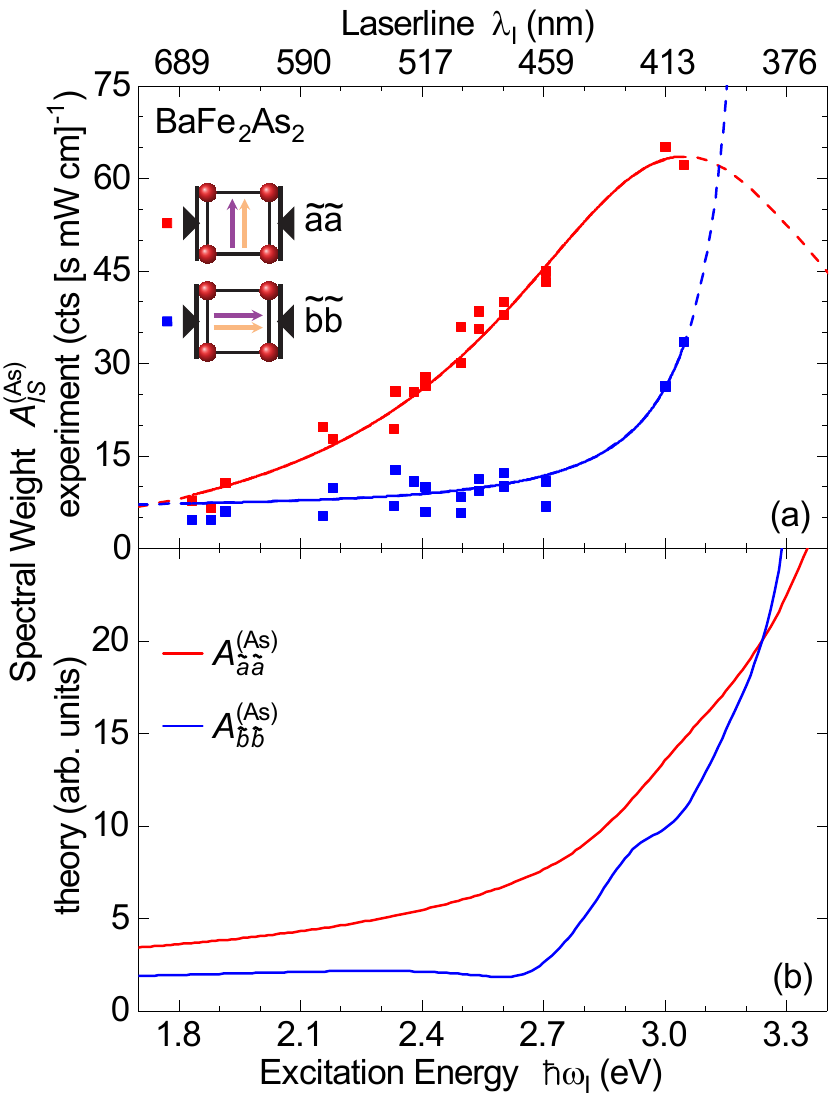}  
\caption{(Color online) Spectral weight $A_{IS}^{\mathrm{(As)}} (\wi)$ of the As phonon as a function of excitation energy and polarization. The top axis shows the corresponding wavelength of the exciting photons. (a) Experimental data. The intensity for parallel light polarizations along the ferromagnetic axis ($\tilde{b}\tilde{b}$, blue squares) is virtually constant for $\hbar\wi < 2.7$\,eV and increases rapidly for $\hbar\wi > 2.7$\,eV. For light polarizations along the antiferromagnetic axis ($\tilde{a}\tilde{a}$, red squares), the phonon intensity increases monotonically over the entire range studied. The solid lines are Lorentzian functions whose extrapolations beyond the measured energy interval are shown as dashed lines. (b) Theoretical prediction of $A^{\mathrm{(As)}}_{\tilde{a}\tilde{a}}$ (red) and $A^{\mathrm{(As)}}_{\tilde{b}\tilde{b}}$ (blue). The curves qualitatively reproduce the experimental data shown in panel (a).}
\label{fig:Ag_resonance_energy}
\end{figure}

We proceed now with the analysis of the phonon spectral weight $A_{IS}^{\mathrm{(As)}}(\wi)$ as a function of the incident photon excitation energy ($\hbar\wi$) and polarization. 
Fig.~\ref{fig:Ag_resonance_energy}(a) shows  $A_{IS}^{\mathrm{(As)}}(\wi)$ as derived by fitting the peak with a Voigt function, after subtracting a linear background. Measurements were repeated several times in order to check the reproducibility. The variation of the spectral weight between different measurements can be taken as an estimate of the experimental error. For light polarizations parallel to  the antiferromagnetic $\tilde{a}$ axis, $A^{\mathrm{(As)}}_{\tilde{a}\tilde{a}}(\wi)$ (red squares) increases continuously with increasing $\hbar\wi$ whereas $A^{\mathrm{(As)}}_{\tilde{b}\tilde{b}}$ (blue squares) stays virtually constant for incident photons in the red and green spectral range, $\hbar\wi < 2.7 \,\mathrm{eV}$, and increases rapidly for $\hbar\wi > 2.7 \,\mathrm{eV}$. For all wavelengths the spectral weight is
higher for the $\tilde{a}\tilde{a}$ than for the $\tilde{b}\tilde{b}$ configuration.

The variations of $A^{\mathrm{(As)}}_{\tilde{a}\tilde{a}}(\wi)$ and $A^{\mathrm{(As)}}_{\tilde{b}\tilde{b}}(\wi)$ display a typical resonance behavior, \cite{Cardona1982_ResonancePhenomena} which is expected when the intermediate state of the Raman scattering process is an eigenstate of the electronic system. Then in second order perturbation theory the intensity diverges as $|\hbar\wi -E_0|^{-2}$ where $E_0$ is the energy difference  between an occupied and an unoccupied electronic Bloch state. In real systems having a finite electronic lifetime a Lorentzian profile is expected. We therefore approximated $A_{IS}^\mathrm{(As)}(\wi)$ with Lorentzians centered at $E_{0,IS}$ as shown by solid lines in Fig.~\ref{fig:Ag_resonance_energy}(a). From these model functions we  determine $E_{0,\tilde{a}\tilde{a}} = 3.1\,\mathrm{eV}$ and $E_{0,\tilde{b}\tilde{b}} = 3.3\,\mathrm{eV}$.

In order to compare the experimental observations with theoretical calculations of the phonon spectral weights the band structure needs to be renormalized so as to account for correlation effects (for details see Appendix~\ref{asec:theory}). 
Specifically, we differentiate three regions: (i)~the unoccupied Fe $3d$ bands near the Fermi energy that we renormalize via a rescaling factor, (ii)~the occupied bands below -2.7\,eV of predominantly As $4p$ character that remain unchanged and (iii)~the occupied bands between -2.7\,eV and the Fermi level derived from hybridized Fe $3d$ and As $4p$ orbitals. 
Due to this hybridization, the renormalization of the latter bands cannot be performed by simple rescaling. 
One can anticipate that the optical absorption would set in at energies below 1.8\,eV, smaller than our minimal laser energy, if the occupied Fe bands would have been renormalized prior to hybridization with the As bands.
Due to the small density of states of the As bands in the range from -2.7\,eV to $E_\mathrm{F}$ their contribution to the dielectric function would be small.
With this in mind, we simply excluded all occupied bands in this range from the calculations.
The effect of these bands, although small, could be accounted for using the DMFT method, which, however, is beyond the scope of our present work.

We then determine the dielectric tensor and the Raman tensor (section~\ref{sec:theory}) on the basis of this renormalized band structure for the $(\pi,0)$ ordered state. Since our resonances lie in the range $\hbar \wi >2.7$\,eV our calculations can capture the intensities and the $\tilde{a}-\tilde{b}$ anisotropy in this range of energies rather well as can be seen in Fig.~\ref{fig:Ag_resonance_energy}(b). 

With this analysis, we interpret our joint experimental and theoretical results as evidence that resonance effects are the main source of the anomalous intensity of the As phonon in crossed polarizations. The main experimental argument is based on the anisotropic variation of the phonon intensities with $\hbar\wi$ in $\tilde{a}\tilde{a}$ and $\tilde{b}\tilde{b}$ polarization configurations, while the theoretical derivation of the tensor elements demonstrates the importance of magnetic order and  correlation effects for the band reconstruction.

As proposed previously,\cite{Garcia:2013} magnetism appears to be the origin of the anisotropy. 
However, the intensity anisotropy cannot be explained without taking into account the high-energy electronic states.

Finally, we briefly looked into the effect of doping on the anomaly and found
further support for its magnetic origin. In \BFCA the transition temperature \TSDW is several degrees  below \Ts for finite $x$, and one observes that the anomaly of the As phonon does not commence at \Ts, but rather at the magnetic transition. For $x=0.025$ the phonon assumes intensity in crossed polarizations only below \TSDW (see supplementary information of Ref.~\onlinecite{Kretzschmar:2016}). For $x=0.051$ the anomaly starts to appear at \Ts, as displayed in Fig.~\ref{Afig:BFCA} in the Appendix, but the spectral weight does not show an order-parameter-like temperature dependence. The increase is nearly linear and saturates below \TSDW at a value which is smaller by a factor of approximately 7 than that in $RR$ polarization projecting \Ag/\Alg symmetry. In FeSe, with a structural transition at $\Ts=89.1$\,K but no long range magnetism, \cite{Baek:2014} the anomalous intensity can also be observed below \Ts but the intensity relative to that in the \Ag projection is only 1\%, as shown in Fig.~\ref{Afig:FeSe}. Similar to Ba(Fe$_{0.949}$Co$_{0.051}$)$_2$As$_2$, the spectral weight increases approximately linearly but does not saturate, presumably because FeSe does not develop long-ranged magnetic order.

\section{Conclusion}

\label{sec:conclusion}

We studied the two \Eg phonons and the fully symmetric As vibration in twin-free \BFA by Raman scattering.
The tetragonal \Eg phonon at 130\,\wn ($\Eg^{(1)}$) splits into two modes in the orthorhombic phase. The detwinning allows us to identify the modes at 125\,\wn and 135\,\wn as \BZg and \BEg phonons, respectively. DFT calculations predict the symmetries correctly and show that the splitting occurs because of the stripe magnetic order.

The As \Ag phonon was studied for various laser lines in the range 1.8 to 3.1\,eV. In the ordered phase the spectral weight of the phonon resonates for an excitation energy of ($3.2 \pm 0.1$)\,eV. The resonance energy is almost the same for the light polarized along the ferro- or
antiferromagnetic directions $\tilde{b}$ and $\tilde{a}$ [for the definition of the axes see Fig.~\ref{fig:orth_phase}(a)], whereas the variation of the spectral weight with the energy of the incident photon is rather different for the $\tilde{b}\tilde{b}$ and $\tilde{a}\tilde{a}$  configurations.

We find that our DFT calculations reproduce the anisotropy and the resonance very well for energies above 2.7\,eV if we include both the effects of the magnetism and of the strong correlations in the Fe $3d$ orbitals, responsible for the band renormalization. For energies below 2.7\,eV our approximation is only semi-quantitative, because the occupied Fe $3d$ bands are strongly hybridized with the As bands and cannot be renormalized by simple rescaling. As in the case of the \Eg phonons, all effects are strongly linked to magnetism. However, in the case of the As phonon the inclusion of electronic states at high energies is essential because of the
resonance behavior. Therefore, low-energy physics with magnetism-induced anisotropic electron-phonon coupling \cite{Garcia:2013} is probably insufficient for explaining the anomalous intensity in crossed polarizations.

\section*{Acknowledgement}

We gratefully acknowledge discussions with L. Degiorgi and thank him for providing us with raw and analyzed IR data of \BFA. 
The work was supported by the German Research Foundation (DFG) via the Priority Program SPP\,1458, the Transregional Collaborative Research Centers TRR\,80, TRR\,49 and by the Serbian Ministry of Education, Science and Technological Development under Project III45018. 
We acknowledge support by the DAAD through the bilateral project between Serbia and Germany (grant numbers 56267076 and 57142964).
The collaboration with Stanford University was supported by the Bavaria California Technology Center BaCaTeC (grant-no. A5\,[2012-2]). Work in the SIMES at Stanford University and SLAC was supported by the U.S. Department of Energy, Office of Basic Energy Sciences, Division of Materials Sciences and Engineering, under Contract No. DE-AC02-76SF00515. 
Y.L. and R.V. acknowledge the allotment of computer time at the Centre for Scientific Computing (CSC) in Frankfurt. 
I.I.M. was supported by ONR through the NRL basic research program and by the Alexander von Humboldt foundation.

\clearpage
\begin{appendix}
\label{sec:appendix}

\section{Calibration of the sensitivity}
\label{asec:dataevaluation}

Scattering experiments performed over a wide energy range necessitate an appropriate correction of the data.
The quantity of interest is the response function $R\chi^{\prime\prime}_{IS}(\Omega)$ where $\Omega=\wi -\ws$ is the Raman shift, and \ws is the energy of the scattered photons. $R$ includes all experimental constants and units in a way that $R\chi^{\prime\prime}_{IS}(\Omega)$ is as close as possible to the count rate $\dot{N}^\ast_{I,S}$, measured for a given laser power $P_\mathrm{I}=I_\mathrm{I}\hbar\wi$ absorbed by the sample. $I_\mathrm{I}$ is the number of incoming photons per unit time and $I,S$ refer to both photon energies and polarizations. With $A_\mathrm{f}$ the (nearly) energy-independent area of the laser focus the cross section is given by \cite{Muschler:2010a}
\begin{equation} \label{eq:cts}
 \frac{\dot{N}^\ast_{IS}(\Delta\ws,\Delta\tilde{\Omega})}{P_I}
 \hbar\wi A_\mathrm{f} =  R^\ast r(\ws)\frac{d^2\sigma}{d\ws d\tilde\Omega}
 \Delta\ws \Delta\tilde{\Omega}.
\end{equation}
$R^\ast$ and $r(\ws)$ are a constant and the relative sensitivity, respectively. $r(\ws)$ is assumed to be dimensionless and includes energy-dependent factors such as surface losses, penetration depth, and the monochromatic efficiency of the setup. $\Delta\ws$ and $\Delta\tilde{\Omega}$ are the bandwidth and the solid angle of acceptance, respectively, and depend both on $\ws$. $r(\ws) \Delta \ws \Delta \tilde{\Omega}$ is determined by calibration and used for correcting the raw data. The resulting rate $\dot{N}_{IS}$ is close to $\dot{N}^\ast_{IS}$ in the range $\Omega \le 1,000$\,cm$^{-1}$ but increasingly different for larger energy transfers mainly for the strong variation of $\Delta\ws$.

Applying the fluctuation-dissipation theorem, one obtains
\begin{eqnarray}
  \frac{\dot{N}_{IS}}{P_I}\hbar\wi A_\mathrm{f}
  &=& R^\prime\frac{d^2\sigma}{d\ws d\tilde\Omega}\nonumber\\
  \label{eq:fd}
  &=& R^\prime\frac{\hbar}{\pi} r_0^2\frac{\ws}{\wi}\{1+n(\Omega,T)\}
  \chi^{\prime\prime}(\Omega),
\end{eqnarray}
where $R^\prime$ is another constant, which is proportional to $\Delta \ws (\omega_0) \Delta \tilde{\Omega}(\omega_0)$, $n(\Omega,T) = [\exp(\frac{\hbar\Omega}{k_BT}) - 1]^{-1}$ is the thermal Bose factor and $r_0$ is the classical electron radius. Finally, after collecting all energy-independent factors in $R$ we obtain
\begin{equation}\label{eq:Rchi}
 R\chi^{\prime\prime}_{IS}(\Omega) = \frac{\dot{N}_{IS}}{P_\mathrm{I}}
 \frac{\wi^2}{\omega_0\ws}
 \left\{1-\exp\left(-\frac{\hbar\Omega}{k_BT}\right)\right\}.
\end{equation}
Here, $\omega_0 = 20,000\,\mathrm{cm}^{-1}$ is inserted for convenience to get a correction close to unity.
Therefore, the spectra shown reflect the measured number of photon counts per second and mW absorbed power as closely as possible, thus approximately obeying counting statistics as intended. Since the spectra are taken with constant slit width the spectral resolution depends on energy, and narrow structures such as phonons may change their shapes but the spectral weight is energy independent.

\section{\texorpdfstring{\Ag}{Ag} spectra}
\label{asec:rawdata}

\begin{figure}
  \centering
  \includegraphics[width=8.5cm]{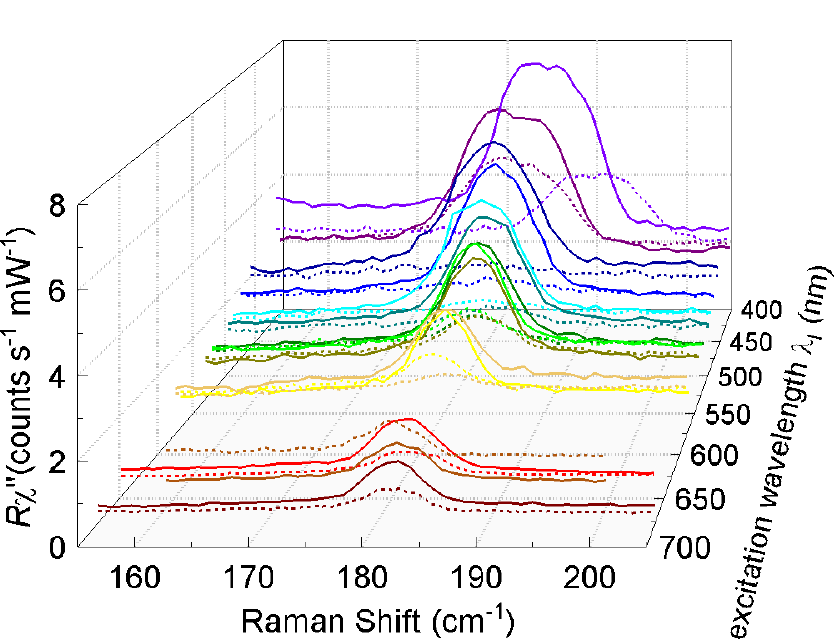}
  \caption[]{(Color online) \Ag spectra of detwinned \BFA at various laser wavelenghts $\lambda_\mathrm{I}$. We used laser lines between 406 and 676\,nm and parallel polarizations of incoming and outgoing photons along the antiferromagnetic ($\tilde{a}\tilde{a}$, solid lines) and the ferromagnetic ($\tilde{b}\tilde{b}$, dashed lines) direction.
  }
  \label{Afig:Ag_all}
\end{figure}

Fig.~\ref{Afig:Ag_all} shows the complete set of the \Ag spectra we measured for detwinned \BFA. All spectra were corrected as described in Appendix~\ref{asec:dataevaluation}. For all spectra the same constant width of $550\,\mu$m of the intermediate slit of the spectrometer was used. This results in an energy-dependent resolution varying between approximately 12\,\wn at 24,630\,\wn (3.05\,eV or 406\,nm) and 3\,\wn at 14,793\,\wn (1.83\,eV or 676\,nm). Accordingly, the width of the peak changes as a function of the excitation wavelength and does not reflect the intrinsic line width of the phonon, in particular not for blue photons. The intensity of the peak monotonically increases towards short wavelengths for the $\tilde{a}\tilde{a}$ spectra (solid lines). For light polarized parallel to the ferromagnetic axis ($\tilde{b}\tilde{b}$, dashed lines) the intensity is low for $\lambda_\mathrm{I} > 450\,\mathrm{nm}$, but strongly increases for $\lambda_\mathrm{I} < 450\,\mathrm{nm}$.  The underlying electronic continuum, which is not a subject of this paper, also changes in intensity as a function of the excitation wavelength.

From the spectra the spectral weight $A_{IS}^{\rm (As)} (\wi)$ of the phonon can be derived by fitting a Voigt function to the phonon peak after subtracting a linear background. The width of the Gaussian part of the Voigt function is given by the known resolution of the spectrometer while that of the Lorentzian part reflects the line width of the phonon.

\section{Spectral weight for \texorpdfstring{$aa$}{aa} and \texorpdfstring{$ab$}{ab} polarizations}
\label{asec:complex_tensor}

\begin{figure}
  \centering
  \includegraphics[width=8.5cm]{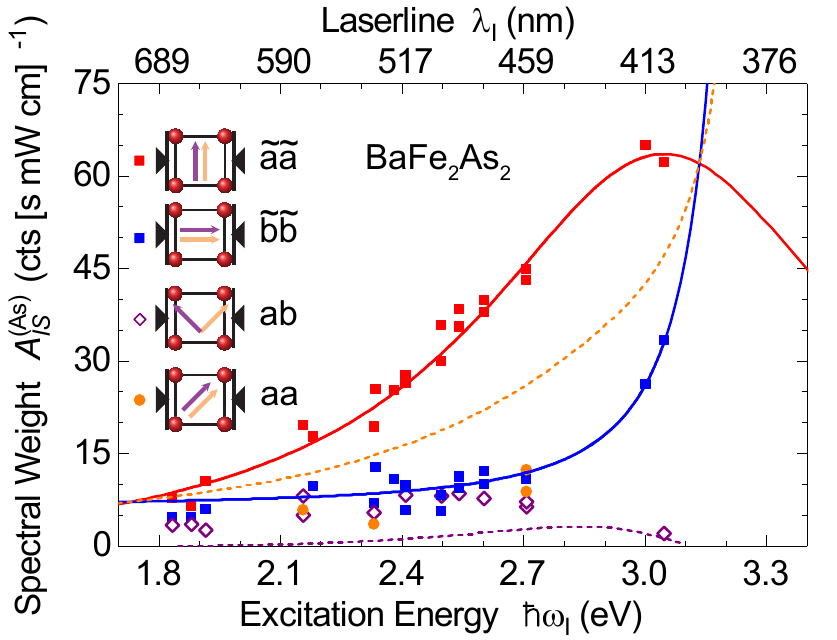}
  \caption{(Color online) Spectral weight $A_{IS}^{\rm (As)} (\wi)$ of the As phonon as a function of excitation energy and polarization. The top axis shows the corresponding wavelength of the exciting photons. The data for $\tilde{a}\tilde{a}$ (red squares) and $\tilde{b}\tilde{b}$ (blue squares) polarizations as well as the Lorentzian model functions (red and blue solid lines) are identical to Fig.~\ref{fig:Ag_resonance_energy}(a). The intensity for crossed ($ab$, purple diamonds) and for $aa$ polarizations (orange dots) is comparable to the intensity found for $\tilde{b}\tilde{b}$ polarization. The purple dashed line is the intensity for $ab$ polarization calculated from the fitted resonance profiles (solid lines) assuming a Raman tensor with real elements. The orange dashed line shows the same calculation for $aa$ polarization.
  }
  \label{afig:Ag_resonance_all_pol}
\end{figure}

For clarity, Fig.~\ref{fig:Ag_resonance_energy}(a) displays only part of the data we collected. We also measured spectra in $aa$ and $ab$ configurations (cf. Figs.~\ref{fig:orth_phase}(a) and \ref{afig:Ag_resonance_all_pol} for the definitions) and find them instructive for various reasons. The $aa$ and $ab$ data (i)~can be compared directly with results presented recently \cite{Wu:2017dec} and (ii)~indicate that the Raman tensor has large imaginary parts.

Fig.~\ref{afig:Ag_resonance_all_pol} shows the spectral weights of the As phonon mode for $aa$ and $ab$ polarizations, $A_{aa}^\mathrm{(As)}(\wi)$ (orange circles) and $A_{ab}^\mathrm{(As)}(\wi)$ (open purple diamonds), respectively, for selected wavelengths together with the data and model functions from Fig.~\ref{fig:Ag_resonance_energy}(a) of the main text. Given the experimental error the respective intensities for $aa$ and $ab$ polarizations are rather similar and are also comparable to $A^{\mathrm(As)}_{\tilde{b}\tilde{b}}(\wi)$ (blue squares) in the range $1.9 < \hbar\wi \le 2.7$\,eV. For $\hbar\wi =3.05$\,eV (406\,nm) $A^{\mathrm(As)}_{ab}(\wi)$ is very small, for the yellow-green spectral range $A^{\mathrm(As)}_{ab}(\wi)$ may be even larger than $A^{\mathrm(As)}_{aa}(\wi)$ in qualitative agreement with Ref.~\onlinecite{Wu:2017dec}.

The elements of a real Raman tensor $\hat{\alpha}^{(\rm Ag)}$ (Eq.~\eqref{eq:tensor-Ag} of the main text) can be derived directly from the experimental data as ${\alpha}_{11} = \sqrt{A^{\mathrm(As)}_{\tilde{a}\tilde{a}}}$ and ${\alpha}_{22} = \sqrt{A^{\mathrm(As)}_{\tilde{b}\tilde{b}}}$. Then, the phonon's spectral weight expected for all other polarizations can be calculated right away, and $A^{\mathrm(As)}_{{a}{a}}$ is just the average of $A^{\mathrm(As)}_{\tilde{a}\tilde{a}}$ and $A^{\mathrm(As)}_{\tilde{b}\tilde{b}}$ (dashed orange line in Fig.~\ref{afig:Ag_resonance_all_pol}).
Obviously, there is no agreement with the experimental values for $A^{\mathrm(As)}_{aa}$ (orange circles).

$A^{\mathrm(As)}_{ab}$ can be determined in a similar fashion. In Fig.~\ref{afig:Ag_resonance_all_pol} we show the expected spectral weight as purple dashed line. The dependence on \wi is again derived from the model functions describing the resonance (full red and blue lines). Also for $A^{\mathrm(As)}_{ab}$ the mismatch between experiment (open purple diamonds) and expectation (purple dashed line) is statistically significant, and one has to conclude that the assumption of real tensor elements in the orthorhombic phase is not valid.

This effect is not particularly surprising in an absorbing material and was in fact discussed earlier for the cuprates. \cite{Strach:1998,Ambrosch:2002} For the Fe-based systems, the possibility of complex Raman tensor elements for the As phonon was not considered yet. Our experimental observations show that the complex nature of $\hat{\alpha}^{\rm (Ag)}$ is crucially important and that the imaginary parts of ${\alpha}_{11}$ and ${\alpha}_{22}$ must have opposite sign to explain the observed enhancement of $A_{ab}^\mathrm{(As)}(\wi)$ and the suppression of $A_{aa}^\mathrm{(As)}(\wi)$ with respect to the values expected for real tensor elements (dashed orange and purple lines in Fig.~\ref{afig:Ag_resonance_all_pol}).

In summary, the results for $A^{\mathrm(As)}_{aa}(\wi)$ and $A^{\mathrm(As)}_{ab}(\wi)$ support our interpretation that absorption processes are important for the proper interpretation of the Raman data. Currently, we cannot imagine anything else but resonance effects due to interband transitions as the source.

\section{Band structure and PDOS}
\label{asec:theory}

\begin{figure}
  \centering
  \includegraphics[width=8.5cm]{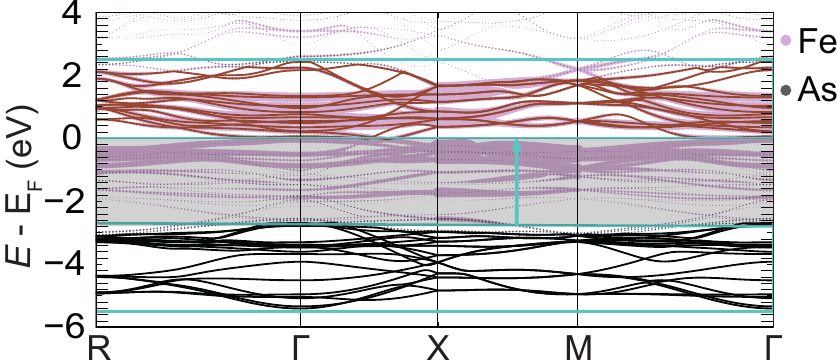}
  \caption{(Color online) DFT band structure. Bands predominantly from Fe states are shown in brown, bands predominantly from As states in black. The shaded region from -2.7\,eV to $E_\mathrm{F}$ contains bands of mixed character and is blacked out for the calculation of the dielectric tensor. Only transitions between the bands within the turquois frames are included.
  }
\label{Afig:bandstructure}
\end{figure}

The DFT band structure is shown in Fig.~\ref{Afig:bandstructure}. Bands above $E_\mathrm{F}$ stem predominantly from Fe $3d$ orbitals (brown) while for $E < -2.7\,\mathrm{eV}$ As $4p$ orbitals prevail (black). 
For a suitable comparison to the experiment these Fe bands are renormalized by a factor between 2 and 3 (Refs.~\onlinecite{Skornyakov:2009,Yao:2011,Ferber:2012,Backes:2015}) while no renormalization is needed for the As bands.
The bands between -2.7\,eV and $E_\mathrm{F}$ are of mixed Fe/As character and are left out when calculating the dielectric tensor as is illustrated by the grey shade in Fig.~\ref{Afig:bandstructure}. Only transitions between the ranges [-5.5\,eV,-2.7\,eV] and [0,2.6\,eV], highlighted by turqouis rectangles, are taken into account.
Thus for photon energies below 2.7\,eV the absorption in our calculations originates predominantly from the Drude response whereas for $\hbar\wi >2.7$\,eV the results become increasingly realistic since they include interband absorption. In either case we use a phenomenological damping of 0.1\,eV. We determine the dielectric tensor and the Raman tensor as described in section \ref{sec:theory} on the basis of this renormalized band structure.
While the $\tilde{a}-\tilde{b}$ anisotropy is qualitatively reproduced for all energies \wi as shown in Fig.~\ref{fig:Ag_resonance_energy}(b) of the main text, the two other experimental quantities, $A_{ab}^\mathrm{(As)}(\wi)$ (purple) and $A_{aa}^\mathrm{(As)}(\wi)$ (orange) shown in Fig.~\ref{afig:Ag_resonance_all_pol} here, are not captured properly simply because the imaginary parts of the theoretically determined tensor elements ${\alpha}^{\mathrm{(As)}\prime\prime}_{ii}$ become very small below 2.7\,eV.
In order to describe $A_{aa}^\mathrm{(As)}(\wi)$ and $A_{ab}^\mathrm{(As)}(\wi)$ absorption processes which lead to imaginary parts of the Raman tensor are necessary.
Upon phenomenologically introducing imaginary parts of $\hat{\alpha}$ for low energies, which cut off at 2.7\,eV where the correct absorption takes over, full agreement can be achieved. However, a solution on a microscopic basis becomes possible only by using LDA$+$DMFT schemes which are beyond the scope of the present work.

\section{\texorpdfstring{\Ag}{Ag} phonon in crossed polarizations in \texorpdfstring{\BFCA}{BaFeCoAs} and FeSe}
\label{asec:Agnematic}
%
\begin{figure}[htb]
  \centering
  \includegraphics[width=8.5cm]{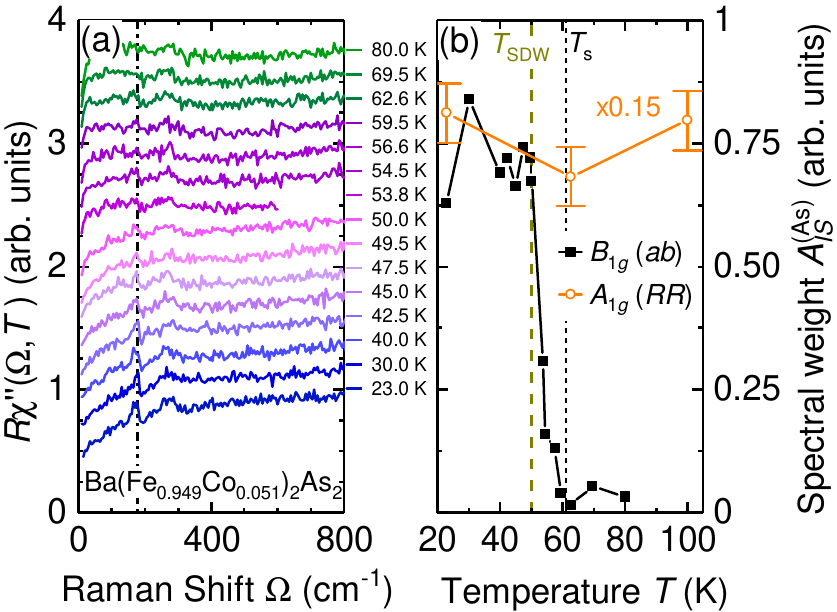}
  \caption{(Color online) As phonon in crossed polarizations for \BFCA at $x=0.051$. (a) Raw data for temperatures as indicated. The phonon position is shown as vertical dash-dotted line. The spectra are shifted vertically for clarity. (b) Temperature dependence of the spectral weight. The spectral weight in \Alg symmetry (orange circles) was multiplied by 0.15. \Ts and \TSDW are indicated as vertical dashed lines.%
  }
\label{Afig:BFCA}
\end{figure}

Similarly as in \BFA, the \Ag phonon can be observed for crossed polarizations. Fig.~\ref{Afig:BFCA}(a) shows Raman spectra in $ab$ polarization of \BFCA with $x=5.1$\% having $\Ts = 60.9$\,K and $\TSDW = 50.0$\,K. The As mode appears below \Ts and gains strength upon cooling. Fig.~\ref{Afig:BFCA}(b) shows the corresponding spectral weight as a function of temperature. In the nematic phase $\TSDW < T < \Ts$ the phonon spectral weight increases almost linearly upon cooling and saturates in the magnetic phase for $T < \TSDW$ at approximately 15\% of that in the fully symmetric channel (\Ag/\Alg).

\begin{figure}
  \centering
  \includegraphics[width=8.5cm]{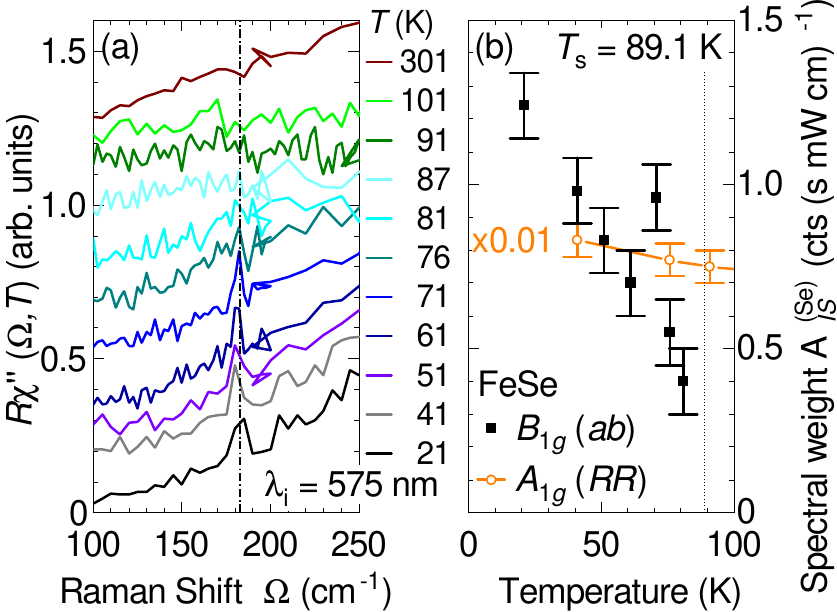}
  \caption{(Color online) Se phonon in crossed polarization for FeSe. (a) Raw data for temperatures as indicated. The phonon position is shown as dash-dotted line. The spectra are shifted vertically for clarity. (b) Temperature dependence of the spectral weight. The spectral weight in \Alg symmetry (orange circles) was multiplied by 0.01. \Ts is indicated as vertical dotted line.%
  }
\label{Afig:FeSe}
\end{figure}

In FeSe the Se phonon appears also in the $ab$ spectra as shown in Fig.~\ref{Afig:FeSe}(a) when the temperature is lowered below the structural phase transition at $\Ts \approx 90$\,K. Upon cooling [Fig.~\ref{Afig:FeSe}(b)] the spectral weight of the phonon increases almost linearly for crossed polarizations ($ab$, black squares), but stays virtually constant across the phase transition for parallel light polarizations [$RR$, orange circles in Fig.~\ref{Afig:FeSe}(b)]. As opposed to \BFCA no saturation of the spectral weight in $ab$ polarizations is found, likely because FeSe shows no long range magnetic order down to lowest temperatures. \cite{Baek:2014} Only about 1\% of the spectral weight of the \Alg spectra ($RR$) is found in crossed polarizations here, in contrast to \BFA and \BFCA, where the spectral weight of the phonon is larger (Figs.~\ref{fig:Ag_Phonon_in_B1g} and \ref{Afig:BFCA}).

\end{appendix}
\newpage 


\begin{thebibliography}{57}%
\makeatletter
\providecommand \@ifxundefined [1]{%
 \@ifx{#1\undefined}
}%
\providecommand \@ifnum [1]{%
 \ifnum #1\expandafter \@firstoftwo
 \else \expandafter \@secondoftwo
 \fi
}%
\providecommand \@ifx [1]{%
 \ifx #1\expandafter \@firstoftwo
 \else \expandafter \@secondoftwo
 \fi
}%
\providecommand \natexlab [1]{#1}%
\providecommand \enquote  [1]{``#1''}%
\providecommand \bibnamefont  [1]{#1}%
\providecommand \bibfnamefont [1]{#1}%
\providecommand \citenamefont [1]{#1}%
\providecommand \href@noop [0]{\@secondoftwo}%
\providecommand \href [0]{\begingroup \@sanitize@url \@href}%
\providecommand \@href[1]{\@@startlink{#1}\@@href}%
\providecommand \@@href[1]{\endgroup#1\@@endlink}%
\providecommand \@sanitize@url [0]{\catcode `\\12\catcode `\$12\catcode
  `\&12\catcode `\#12\catcode `\^12\catcode `\_12\catcode `\%12\relax}%
\providecommand \@@startlink[1]{}%
\providecommand \@@endlink[0]{}%
\providecommand \url  [0]{\begingroup\@sanitize@url \@url }%
\providecommand \@url [1]{\endgroup\@href {#1}{\urlprefix }}%
\providecommand \urlprefix  [0]{URL }%
\providecommand \Eprint [0]{\href }%
\providecommand \doibase [0]{http://dx.doi.org/}%
\providecommand \selectlanguage [0]{\@gobble}%
\providecommand \bibinfo  [0]{\@secondoftwo}%
\providecommand \bibfield  [0]{\@secondoftwo}%
\providecommand \translation [1]{[#1]}%
\providecommand \BibitemOpen [0]{}%
\providecommand \bibitemStop [0]{}%
\providecommand \bibitemNoStop [0]{.\EOS\space}%
\providecommand \EOS [0]{\spacefactor3000\relax}%
\providecommand \BibitemShut  [1]{\csname bibitem#1\endcsname}%
\let\auto@bib@innerbib\@empty
\bibitem [{\citenamefont {Sefat}\ and\ \citenamefont
  {Singh}(2011)}]{Sefat:2011}%
  \BibitemOpen
  \bibfield  {author} {\bibinfo {author} {\bibfnamefont {Athena~S.}\
  \bibnamefont {Sefat}}\ and\ \bibinfo {author} {\bibfnamefont {David~J.}\
  \bibnamefont {Singh}},\ }\bibfield  {title} {\enquote {\bibinfo {title}
  {{Chemistry and electronic structure of iron-based superconductors}},}\
  }\href {\doibase 10.1557/mrs.2011.175} {\bibfield  {journal} {\bibinfo
  {journal} {MRS Bull.}\ }\textbf {\bibinfo {volume} {36}},\ \bibinfo {pages}
  {614--619} (\bibinfo {year} {2011})}\BibitemShut {NoStop}%
\bibitem [{\citenamefont {Wang}\ \emph {et~al.}(2015)\citenamefont {Wang},
  \citenamefont {Kivelson},\ and\ \citenamefont {Lee}}]{Wang:2015}%
  \BibitemOpen
  \bibfield  {author} {\bibinfo {author} {\bibfnamefont {Fa}~\bibnamefont
  {Wang}}, \bibinfo {author} {\bibfnamefont {Steven~A.}\ \bibnamefont
  {Kivelson}}, \ and\ \bibinfo {author} {\bibfnamefont {Dung-Hai}\ \bibnamefont
  {Lee}},\ }\bibfield  {title} {\enquote {\bibinfo {title} {{Nematicity and
  quantum paramagnetism in FeSe}},}\ }\href {\doibase 10.1038/nphys3456}
  {\bibfield  {journal} {\bibinfo  {journal} {Nat. Phys.}\ }\textbf {\bibinfo
  {volume} {11}},\ \bibinfo {pages} {959--963} (\bibinfo {year}
  {2015})}\BibitemShut {NoStop}%
\bibitem [{\citenamefont {Gallais}\ and\ \citenamefont
  {Paul}(2016)}]{Gallais:2016a}%
  \BibitemOpen
  \bibfield  {author} {\bibinfo {author} {\bibfnamefont {Yann}\ \bibnamefont
  {Gallais}}\ and\ \bibinfo {author} {\bibfnamefont {Indranil}\ \bibnamefont
  {Paul}},\ }\bibfield  {title} {\enquote {\bibinfo {title} {{Charge nematicity
  and electronic Raman scattering in iron-based superconductors}},}\ }\href
  {\doibase http://dx.doi.org/10.1016/j.crhy.2015.10.001} {\bibfield  {journal}
  {\bibinfo  {journal} {C. R. Phys.}\ }\textbf {\bibinfo {volume} {17}},\
  \bibinfo {pages} {113--139} (\bibinfo {year} {2016})}\BibitemShut {NoStop}%
\bibitem [{\citenamefont {Yi}\ \emph {et~al.}(2017)\citenamefont {Yi},
  \citenamefont {Zhang}, \citenamefont {Shen},\ and\ \citenamefont
  {Lu}}]{YiM:2017}%
  \BibitemOpen
  \bibfield  {author} {\bibinfo {author} {\bibfnamefont {Ming}\ \bibnamefont
  {Yi}}, \bibinfo {author} {\bibfnamefont {Yan}\ \bibnamefont {Zhang}},
  \bibinfo {author} {\bibfnamefont {Zhi-Xun}\ \bibnamefont {Shen}}, \ and\
  \bibinfo {author} {\bibfnamefont {Donghui}\ \bibnamefont {Lu}},\ }\bibfield
  {title} {\enquote {\bibinfo {title} {{Role of the orbital degree of freedom
  in iron-based superconductors}},}\ }\href {\doibase
  10.1038/s41535-017-0059-y} {\bibfield  {journal} {\bibinfo  {journal} {npj
  Quantum Mater.}\ }\textbf {\bibinfo {volume} {2}},\ \bibinfo {pages} {57}
  (\bibinfo {year} {2017})}\BibitemShut {NoStop}%
\bibitem [{\citenamefont {B\"{o}hmer}\ and\ \citenamefont
  {Kreisel}(2018)}]{Bohmer:2018}%
  \BibitemOpen
  \bibfield  {author} {\bibinfo {author} {\bibfnamefont {Anna~E.}\ \bibnamefont
  {B\"{o}hmer}}\ and\ \bibinfo {author} {\bibfnamefont {Andreas}\ \bibnamefont
  {Kreisel}},\ }\bibfield  {title} {\enquote {\bibinfo {title} {{Nematicity,
  magnetism and superconductivity in FeSe}},}\ }\href {\doibase
  10.1088/1361-648X/aa9caa} {\bibfield  {journal} {\bibinfo  {journal} {J.
  Phys.: Condens. Matter}\ }\textbf {\bibinfo {volume} {30}},\ \bibinfo {pages}
  {023001} (\bibinfo {year} {2018})}\BibitemShut {NoStop}%
\bibitem [{\citenamefont {Yildirim}(2009)}]{Yildirim:2009a}%
  \BibitemOpen
  \bibfield  {author} {\bibinfo {author} {\bibfnamefont {T.}~\bibnamefont
  {Yildirim}},\ }\bibfield  {title} {\enquote {\bibinfo {title} {{Strong
  Coupling of the Fe-Spin State and the As-As Hybridization in Iron-Pnictide
  Superconductors from First-Principle Calculations}},}\ }\href {\doibase
  10.1103/PhysRevLett.102.037003} {\bibfield  {journal} {\bibinfo  {journal}
  {Phys. Rev. Lett.}\ }\textbf {\bibinfo {volume} {102}},\ \bibinfo {pages}
  {037003} (\bibinfo {year} {2009})}\BibitemShut {NoStop}%
\bibitem [{\citenamefont {Zbiri}\ \emph {et~al.}(2009)\citenamefont {Zbiri},
  \citenamefont {Schober}, \citenamefont {Johnson}, \citenamefont {Rols},
  \citenamefont {Mittal}, \citenamefont {Su}, \citenamefont {Rotter},\ and\
  \citenamefont {Johrendt}}]{Zbiri:2009i}%
  \BibitemOpen
  \bibfield  {author} {\bibinfo {author} {\bibfnamefont {M.}~\bibnamefont
  {Zbiri}}, \bibinfo {author} {\bibfnamefont {H.}~\bibnamefont {Schober}},
  \bibinfo {author} {\bibfnamefont {M.~R.}\ \bibnamefont {Johnson}}, \bibinfo
  {author} {\bibfnamefont {S.}~\bibnamefont {Rols}}, \bibinfo {author}
  {\bibfnamefont {R.}~\bibnamefont {Mittal}}, \bibinfo {author} {\bibfnamefont
  {Y.}~\bibnamefont {Su}}, \bibinfo {author} {\bibfnamefont {M.}~\bibnamefont
  {Rotter}}, \ and\ \bibinfo {author} {\bibfnamefont {D.}~\bibnamefont
  {Johrendt}},\ }\bibfield  {title} {\enquote {\bibinfo {title} {{Ab initio
  lattice dynamics simulations and inelastic neutron scattering spectra for
  studying phonons in BaFe$_{2}$As$_{2}$: Effect of structural phase
  transition, structural relaxation, and magnetic ordering}},}\ }\href
  {\doibase 10.1103/PhysRevB.79.064511} {\bibfield  {journal} {\bibinfo
  {journal} {Phys. Rev. B}\ }\textbf {\bibinfo {volume} {79}},\ \bibinfo
  {pages} {064511} (\bibinfo {year} {2009})}\BibitemShut {NoStop}%
\bibitem [{\citenamefont {Chauvi\`ere}\ \emph {et~al.}(2009)\citenamefont
  {Chauvi\`ere}, \citenamefont {Gallais}, \citenamefont {Cazayous},
  \citenamefont {Sacuto}, \citenamefont {M\'easson}, \citenamefont {Colson},\
  and\ \citenamefont {Forget}}]{Chauviere:2009}%
  \BibitemOpen
  \bibfield  {author} {\bibinfo {author} {\bibfnamefont {L.}~\bibnamefont
  {Chauvi\`ere}}, \bibinfo {author} {\bibfnamefont {Y.}~\bibnamefont
  {Gallais}}, \bibinfo {author} {\bibfnamefont {M.}~\bibnamefont {Cazayous}},
  \bibinfo {author} {\bibfnamefont {A.}~\bibnamefont {Sacuto}}, \bibinfo
  {author} {\bibfnamefont {M.~A.}\ \bibnamefont {M\'easson}}, \bibinfo {author}
  {\bibfnamefont {D.}~\bibnamefont {Colson}}, \ and\ \bibinfo {author}
  {\bibfnamefont {A.}~\bibnamefont {Forget}},\ }\bibfield  {title} {\enquote
  {\bibinfo {title} {{Doping dependence of the lattice dynamics in Ba(Fe$_{1 -
  x}$Co$_x$)$_2$As$_2$ studied by Raman spectroscopy}},}\ }\href {\doibase
  10.1103/PhysRevB.80.094504} {\bibfield  {journal} {\bibinfo  {journal} {Phys.
  Rev. B}\ }\textbf {\bibinfo {volume} {80}},\ \bibinfo {eid} {094504}
  (\bibinfo {year} {2009})}\BibitemShut {NoStop}%
\bibitem [{\citenamefont {Chauvi\`ere}\ \emph {et~al.}(2011)\citenamefont
  {Chauvi\`ere}, \citenamefont {Gallais}, \citenamefont {Cazayous},
  \citenamefont {M\'easson}, \citenamefont {Sacuto}, \citenamefont {Colson},\
  and\ \citenamefont {Forget}}]{Chauviere:2011}%
  \BibitemOpen
  \bibfield  {author} {\bibinfo {author} {\bibfnamefont {L.}~\bibnamefont
  {Chauvi\`ere}}, \bibinfo {author} {\bibfnamefont {Y.}~\bibnamefont
  {Gallais}}, \bibinfo {author} {\bibfnamefont {M.}~\bibnamefont {Cazayous}},
  \bibinfo {author} {\bibfnamefont {M.~A.}\ \bibnamefont {M\'easson}}, \bibinfo
  {author} {\bibfnamefont {A.}~\bibnamefont {Sacuto}}, \bibinfo {author}
  {\bibfnamefont {D.}~\bibnamefont {Colson}}, \ and\ \bibinfo {author}
  {\bibfnamefont {A.}~\bibnamefont {Forget}},\ }\bibfield  {title} {\enquote
  {\bibinfo {title} {{Raman scattering study of spin-density-wave order and
  electron-phonon coupling in Ba(Fe$_{1-x}$Co$_{x}$)$_{2}$As$_{2}$}},}\ }\href
  {\doibase 10.1103/PhysRevB.84.104508} {\bibfield  {journal} {\bibinfo
  {journal} {Phys. Rev. B}\ }\textbf {\bibinfo {volume} {84}},\ \bibinfo
  {pages} {104508} (\bibinfo {year} {2011})}\BibitemShut {NoStop}%
\bibitem [{\citenamefont {Rahlenbeck}\ \emph {et~al.}(2009)\citenamefont
  {Rahlenbeck}, \citenamefont {Sun}, \citenamefont {Sun}, \citenamefont {Lin},
  \citenamefont {Keimer},\ and\ \citenamefont {Ulrich}}]{Rahlenbeck:2009}%
  \BibitemOpen
  \bibfield  {author} {\bibinfo {author} {\bibfnamefont {M.}~\bibnamefont
  {Rahlenbeck}}, \bibinfo {author} {\bibfnamefont {G.~L.}\ \bibnamefont {Sun}},
  \bibinfo {author} {\bibfnamefont {D.~L.}\ \bibnamefont {Sun}}, \bibinfo
  {author} {\bibfnamefont {C.~T.}\ \bibnamefont {Lin}}, \bibinfo {author}
  {\bibfnamefont {B.}~\bibnamefont {Keimer}}, \ and\ \bibinfo {author}
  {\bibfnamefont {C.}~\bibnamefont {Ulrich}},\ }\bibfield  {title} {\enquote
  {\bibinfo {title} {{Phonon anomalies in pure and underdoped R$_{1 -
  x}$K$_x$Fe$_2$As$_2$ (R = Ba, Sr) investigated by Raman light scattering}},}\
  }\href {\doibase 10.1103/PhysRevB.80.064509} {\bibfield  {journal} {\bibinfo
  {journal} {Phys. Rev. B}\ }\textbf {\bibinfo {volume} {80}},\ \bibinfo
  {pages} {064509} (\bibinfo {year} {2009})}\BibitemShut {NoStop}%
\bibitem [{\citenamefont {Kumar}\ \emph {et~al.}(2010)\citenamefont {Kumar},
  \citenamefont {Kumar}, \citenamefont {Saha}, \citenamefont {Muthu},
  \citenamefont {Prakash}, \citenamefont {Patnaik}, \citenamefont {Waghmare},
  \citenamefont {Ganguli},\ and\ \citenamefont {Sood}}]{Kumar:2010}%
  \BibitemOpen
  \bibfield  {author} {\bibinfo {author} {\bibfnamefont {Pradeep}\ \bibnamefont
  {Kumar}}, \bibinfo {author} {\bibfnamefont {Anil}\ \bibnamefont {Kumar}},
  \bibinfo {author} {\bibfnamefont {Surajit}\ \bibnamefont {Saha}}, \bibinfo
  {author} {\bibfnamefont {D.V.S.}\ \bibnamefont {Muthu}}, \bibinfo {author}
  {\bibfnamefont {J.}~\bibnamefont {Prakash}}, \bibinfo {author} {\bibfnamefont
  {S.}~\bibnamefont {Patnaik}}, \bibinfo {author} {\bibfnamefont {U.V.}\
  \bibnamefont {Waghmare}}, \bibinfo {author} {\bibfnamefont {A.K.}\
  \bibnamefont {Ganguli}}, \ and\ \bibinfo {author} {\bibfnamefont {A.K.}\
  \bibnamefont {Sood}},\ }\bibfield  {title} {\enquote {\bibinfo {title}
  {{Anomalous Raman scattering from phonons and electrons of superconducting
  FeSe$_{0.82}$}},}\ }\href {\doibase 10.1016/j.ssc.2010.01.033} {\bibfield
  {journal} {\bibinfo  {journal} {Solid State Commun.}\ }\textbf {\bibinfo
  {volume} {150}},\ \bibinfo {pages} {557} (\bibinfo {year}
  {2010})}\BibitemShut {NoStop}%
\bibitem [{\citenamefont {Mittal}\ \emph {et~al.}(2009)\citenamefont {Mittal},
  \citenamefont {Pintschovius}, \citenamefont {Lamago}, \citenamefont {Heid},
  \citenamefont {Bohnen}, \citenamefont {Reznik}, \citenamefont {Chaplot},
  \citenamefont {Su}, \citenamefont {Kumar}, \citenamefont {Dhar},
  \citenamefont {Thamizhavel},\ and\ \citenamefont {Brueckel}}]{Mittal:2009}%
  \BibitemOpen
  \bibfield  {author} {\bibinfo {author} {\bibfnamefont {R.}~\bibnamefont
  {Mittal}}, \bibinfo {author} {\bibfnamefont {L.}~\bibnamefont
  {Pintschovius}}, \bibinfo {author} {\bibfnamefont {D.}~\bibnamefont
  {Lamago}}, \bibinfo {author} {\bibfnamefont {R.}~\bibnamefont {Heid}},
  \bibinfo {author} {\bibfnamefont {K.-P.}\ \bibnamefont {Bohnen}}, \bibinfo
  {author} {\bibfnamefont {D.}~\bibnamefont {Reznik}}, \bibinfo {author}
  {\bibfnamefont {S.~L.}\ \bibnamefont {Chaplot}}, \bibinfo {author}
  {\bibfnamefont {Y.}~\bibnamefont {Su}}, \bibinfo {author} {\bibfnamefont
  {N.}~\bibnamefont {Kumar}}, \bibinfo {author} {\bibfnamefont {S.~K.}\
  \bibnamefont {Dhar}}, \bibinfo {author} {\bibfnamefont {A.}~\bibnamefont
  {Thamizhavel}}, \ and\ \bibinfo {author} {\bibfnamefont {Th.}\ \bibnamefont
  {Brueckel}},\ }\bibfield  {title} {\enquote {\bibinfo {title} {{Measurement
  of Anomalous Phonon Dispersion of CaFe$_2$As$_2$ Single Crystals Using
  Inelastic Neutron Scattering}},}\ }\href {\doibase
  10.1103/PhysRevLett.102.217001} {\bibfield  {journal} {\bibinfo  {journal}
  {Phys. Rev. Lett.}\ }\textbf {\bibinfo {volume} {102}},\ \bibinfo {eid}
  {217001} (\bibinfo {year} {2009})}\BibitemShut {NoStop}%
\bibitem [{\citenamefont {Gnezdilov}\ \emph {et~al.}(2013)\citenamefont
  {Gnezdilov}, \citenamefont {Pashkevich}, \citenamefont {Lemmens},
  \citenamefont {Wulferding}, \citenamefont {Shevtsova}, \citenamefont {Gusev},
  \citenamefont {Chareev},\ and\ \citenamefont {Vasiliev}}]{Gnezdilov:2013}%
  \BibitemOpen
  \bibfield  {author} {\bibinfo {author} {\bibfnamefont {Vladimir}\
  \bibnamefont {Gnezdilov}}, \bibinfo {author} {\bibfnamefont {Yurii~G.}\
  \bibnamefont {Pashkevich}}, \bibinfo {author} {\bibfnamefont {Peter}\
  \bibnamefont {Lemmens}}, \bibinfo {author} {\bibfnamefont {Dirk}\
  \bibnamefont {Wulferding}}, \bibinfo {author} {\bibfnamefont {Tatiana}\
  \bibnamefont {Shevtsova}}, \bibinfo {author} {\bibfnamefont {Alexander}\
  \bibnamefont {Gusev}}, \bibinfo {author} {\bibfnamefont {Dmitry}\
  \bibnamefont {Chareev}}, \ and\ \bibinfo {author} {\bibfnamefont {Alexander}\
  \bibnamefont {Vasiliev}},\ }\bibfield  {title} {\enquote {\bibinfo {title}
  {{Interplay between lattice and spin states degree of freedom in the FeSe
  superconductor: Dynamic spin state instabilities}},}\ }\href {\doibase
  10.1103/PhysRevB.87.144508} {\bibfield  {journal} {\bibinfo  {journal} {Phys.
  Rev. B}\ }\textbf {\bibinfo {volume} {87}},\ \bibinfo {pages} {144508}
  (\bibinfo {year} {2013})}\BibitemShut {NoStop}%
\bibitem [{\citenamefont {Gnezdilov}\ \emph {et~al.}(2011)\citenamefont
  {Gnezdilov}, \citenamefont {Pashkevich}, \citenamefont {Lemmens},
  \citenamefont {Gusev}, \citenamefont {Lamonova}, \citenamefont {Shevtsova},
  \citenamefont {Vitebskiy}, \citenamefont {Afanasiev}, \citenamefont
  {Gnatchenko}, \citenamefont {Tsurkan}, \citenamefont {Deisenhofer},\ and\
  \citenamefont {Loidl}}]{Gnezdilov:2011}%
  \BibitemOpen
  \bibfield  {author} {\bibinfo {author} {\bibfnamefont {V.}~\bibnamefont
  {Gnezdilov}}, \bibinfo {author} {\bibfnamefont {Yu.}\ \bibnamefont
  {Pashkevich}}, \bibinfo {author} {\bibfnamefont {P.}~\bibnamefont {Lemmens}},
  \bibinfo {author} {\bibfnamefont {A.}~\bibnamefont {Gusev}}, \bibinfo
  {author} {\bibfnamefont {K.}~\bibnamefont {Lamonova}}, \bibinfo {author}
  {\bibfnamefont {T.}~\bibnamefont {Shevtsova}}, \bibinfo {author}
  {\bibfnamefont {I.}~\bibnamefont {Vitebskiy}}, \bibinfo {author}
  {\bibfnamefont {O.}~\bibnamefont {Afanasiev}}, \bibinfo {author}
  {\bibfnamefont {S.}~\bibnamefont {Gnatchenko}}, \bibinfo {author}
  {\bibfnamefont {V.}~\bibnamefont {Tsurkan}}, \bibinfo {author} {\bibfnamefont
  {J.}~\bibnamefont {Deisenhofer}}, \ and\ \bibinfo {author} {\bibfnamefont
  {A.}~\bibnamefont {Loidl}},\ }\bibfield  {title} {\enquote {\bibinfo {title}
  {{Anomalous optical phonons in FeTe chalcogenides: Spin state, magnetic
  order, and lattice anharmonicity}},}\ }\href {\doibase
  10.1103/PhysRevB.83.245127} {\bibfield  {journal} {\bibinfo  {journal} {Phys.
  Rev. B}\ }\textbf {\bibinfo {volume} {83}},\ \bibinfo {pages} {245127}
  (\bibinfo {year} {2011})}\BibitemShut {NoStop}%
\bibitem [{\citenamefont {Akrap}\ \emph {et~al.}(2009)\citenamefont {Akrap},
  \citenamefont {Tu}, \citenamefont {Li}, \citenamefont {Cao}, \citenamefont
  {Xu},\ and\ \citenamefont {Homes}}]{Akrap:2009}%
  \BibitemOpen
  \bibfield  {author} {\bibinfo {author} {\bibfnamefont {A.}~\bibnamefont
  {Akrap}}, \bibinfo {author} {\bibfnamefont {J.~J.}\ \bibnamefont {Tu}},
  \bibinfo {author} {\bibfnamefont {L.~J.}\ \bibnamefont {Li}}, \bibinfo
  {author} {\bibfnamefont {G.~H.}\ \bibnamefont {Cao}}, \bibinfo {author}
  {\bibfnamefont {Z.~A.}\ \bibnamefont {Xu}}, \ and\ \bibinfo {author}
  {\bibfnamefont {C.~C.}\ \bibnamefont {Homes}},\ }\bibfield  {title} {\enquote
  {\bibinfo {title} {{Infrared phonon anomaly in BaFe$_2$As$_2$}},}\ }\href
  {\doibase 10.1103/PhysRevB.80.180502} {\bibfield  {journal} {\bibinfo
  {journal} {Phys. Rev. B}\ }\textbf {\bibinfo {volume} {80}},\ \bibinfo
  {pages} {180502} (\bibinfo {year} {2009})}\BibitemShut {NoStop}%
\bibitem [{\citenamefont {Garc\'ia-Mart\'inez}\ \emph
  {et~al.}(2013)\citenamefont {Garc\'ia-Mart\'inez}, \citenamefont
  {Valenzuela}, \citenamefont {Ciuchi}, \citenamefont {Cappelluti},
  \citenamefont {Calder\'on},\ and\ \citenamefont {Bascones}}]{Garcia:2013}%
  \BibitemOpen
  \bibfield  {author} {\bibinfo {author} {\bibfnamefont {N.~A.}\ \bibnamefont
  {Garc\'ia-Mart\'inez}}, \bibinfo {author} {\bibfnamefont {B.}~\bibnamefont
  {Valenzuela}}, \bibinfo {author} {\bibfnamefont {S.}~\bibnamefont {Ciuchi}},
  \bibinfo {author} {\bibfnamefont {E.}~\bibnamefont {Cappelluti}}, \bibinfo
  {author} {\bibfnamefont {M.~J.}\ \bibnamefont {Calder\'on}}, \ and\ \bibinfo
  {author} {\bibfnamefont {E.}~\bibnamefont {Bascones}},\ }\bibfield  {title}
  {\enquote {\bibinfo {title} {{Coupling of the As ${A}_{1g}$ phonon to
  magnetism in iron pnictides}},}\ }\href {\doibase 10.1103/PhysRevB.88.165106}
  {\bibfield  {journal} {\bibinfo  {journal} {Phys. Rev. B}\ }\textbf {\bibinfo
  {volume} {88}},\ \bibinfo {pages} {165106} (\bibinfo {year}
  {2013})}\BibitemShut {NoStop}%
\bibitem [{\citenamefont {{Wu}}\ \emph {et~al.}(2017)\citenamefont {{Wu}},
  \citenamefont {{Zhang}}, \citenamefont {{Thorsm{\o}lle}}, \citenamefont
  {{Chen}}, \citenamefont {{Tan}}, \citenamefont {{Dai}}, \citenamefont
  {{Shi}}, \citenamefont {{Jin}}, \citenamefont {{Shibauchi}}, \citenamefont
  {{Kasahara}}, \citenamefont {{Matsuda}}, \citenamefont {{Sefat}},
  \citenamefont {{Ding}}, \citenamefont {{Richard}},\ and\ \citenamefont
  {{Blumberg}}}]{Wu:2017dec}%
  \BibitemOpen
  \bibfield  {author} {\bibinfo {author} {\bibfnamefont {S.-F.}\ \bibnamefont
  {{Wu}}}, \bibinfo {author} {\bibfnamefont {W.-L.}\ \bibnamefont {{Zhang}}},
  \bibinfo {author} {\bibfnamefont {V.~K.}\ \bibnamefont {{Thorsm{\o}lle}}},
  \bibinfo {author} {\bibfnamefont {G.~F.}\ \bibnamefont {{Chen}}}, \bibinfo
  {author} {\bibfnamefont {G.~T.}\ \bibnamefont {{Tan}}}, \bibinfo {author}
  {\bibfnamefont {P.~C.}\ \bibnamefont {{Dai}}}, \bibinfo {author}
  {\bibfnamefont {Y.~G.}\ \bibnamefont {{Shi}}}, \bibinfo {author}
  {\bibfnamefont {C.~Q.}\ \bibnamefont {{Jin}}}, \bibinfo {author}
  {\bibfnamefont {T.}~\bibnamefont {{Shibauchi}}}, \bibinfo {author}
  {\bibfnamefont {S.}~\bibnamefont {{Kasahara}}}, \bibinfo {author}
  {\bibfnamefont {Y.}~\bibnamefont {{Matsuda}}}, \bibinfo {author}
  {\bibfnamefont {A.~S.}\ \bibnamefont {{Sefat}}}, \bibinfo {author}
  {\bibfnamefont {H.}~\bibnamefont {{Ding}}}, \bibinfo {author} {\bibfnamefont
  {P.}~\bibnamefont {{Richard}}}, \ and\ \bibinfo {author} {\bibfnamefont
  {G.}~\bibnamefont {{Blumberg}}},\ }\href@noop {} {\enquote {\bibinfo {title}
  {{Magneto-elastic coupling in Fe-based superconductors}},}\ } (\bibinfo
  {year} {2017}),\ \Eprint {http://arxiv.org/abs/1712.01896} {arXiv:1712.01896
  [cond-mat.supr-con]} \BibitemShut {NoStop}%
\bibitem [{\citenamefont {Mazin}\ \emph {et~al.}(2008)\citenamefont {Mazin},
  \citenamefont {Johannes}, \citenamefont {Boeri}, \citenamefont {Koepernik},\
  and\ \citenamefont {Singh}}]{Mazin:2008a}%
  \BibitemOpen
  \bibfield  {author} {\bibinfo {author} {\bibfnamefont {I.~I.}\ \bibnamefont
  {Mazin}}, \bibinfo {author} {\bibfnamefont {M.~D.}\ \bibnamefont {Johannes}},
  \bibinfo {author} {\bibfnamefont {L.}~\bibnamefont {Boeri}}, \bibinfo
  {author} {\bibfnamefont {K.}~\bibnamefont {Koepernik}}, \ and\ \bibinfo
  {author} {\bibfnamefont {D.~J.}\ \bibnamefont {Singh}},\ }\bibfield  {title}
  {\enquote {\bibinfo {title} {Problems with reconciling density functional
  theory calculations with experiment in ferropnictides},}\ }\href {\doibase
  10.1103/PhysRevB.78.085104} {\bibfield  {journal} {\bibinfo  {journal} {Phys.
  Rev. B}\ }\textbf {\bibinfo {volume} {78}},\ \bibinfo {pages} {085104}
  (\bibinfo {year} {2008})}\BibitemShut {NoStop}%
\bibitem [{\citenamefont {Chu}\ \emph {et~al.}(2009)\citenamefont {Chu},
  \citenamefont {Analytis}, \citenamefont {Kucharczyk},\ and\ \citenamefont
  {Fisher}}]{Chu:2009}%
  \BibitemOpen
  \bibfield  {author} {\bibinfo {author} {\bibfnamefont {Jiun-Haw}\
  \bibnamefont {Chu}}, \bibinfo {author} {\bibfnamefont {James~G.}\
  \bibnamefont {Analytis}}, \bibinfo {author} {\bibfnamefont {Chris}\
  \bibnamefont {Kucharczyk}}, \ and\ \bibinfo {author} {\bibfnamefont {Ian~R.}\
  \bibnamefont {Fisher}},\ }\bibfield  {title} {\enquote {\bibinfo {title}
  {{Determination of the phase diagram of the electron-doped superconductor
  Ba(Fe$_{1 - x}$Co$_x$)$_2$As$_2$}},}\ }\href {\doibase
  10.1103/PhysRevB.79.014506} {\bibfield  {journal} {\bibinfo  {journal} {Phys.
  Rev. B}\ }\textbf {\bibinfo {volume} {79}},\ \bibinfo {eid} {014506}
  (\bibinfo {year} {2009})}\BibitemShut {NoStop}%
\bibitem [{\citenamefont {Kimber}\ \emph {et~al.}(2009)\citenamefont {Kimber},
  \citenamefont {Kreyssig}, \citenamefont {Zhang}, \citenamefont {Jeschke},
  \citenamefont {Valent\'i}, \citenamefont {Yokaichiya}, \citenamefont
  {Colombier}, \citenamefont {Yan}, \citenamefont {Hansen}, \citenamefont
  {Chatterji}, \citenamefont {McQueeney}, \citenamefont {Canfield},
  \citenamefont {Goldman},\ and\ \citenamefont {Argyriou}}]{Kimber:2009i}%
  \BibitemOpen
  \bibfield  {author} {\bibinfo {author} {\bibfnamefont {S.~A.~J.}\
  \bibnamefont {Kimber}}, \bibinfo {author} {\bibfnamefont {A.}~\bibnamefont
  {Kreyssig}}, \bibinfo {author} {\bibfnamefont {Yu-Zhong}\ \bibnamefont
  {Zhang}}, \bibinfo {author} {\bibfnamefont {H.~O.}\ \bibnamefont {Jeschke}},
  \bibinfo {author} {\bibfnamefont {R.}~\bibnamefont {Valent\'i}}, \bibinfo
  {author} {\bibfnamefont {F.}~\bibnamefont {Yokaichiya}}, \bibinfo {author}
  {\bibfnamefont {E.}~\bibnamefont {Colombier}}, \bibinfo {author}
  {\bibfnamefont {Jiaqiang}\ \bibnamefont {Yan}}, \bibinfo {author}
  {\bibfnamefont {T.~C.}\ \bibnamefont {Hansen}}, \bibinfo {author}
  {\bibfnamefont {T.}~\bibnamefont {Chatterji}}, \bibinfo {author}
  {\bibfnamefont {R.~J.}\ \bibnamefont {McQueeney}}, \bibinfo {author}
  {\bibfnamefont {P.~C.}\ \bibnamefont {Canfield}}, \bibinfo {author}
  {\bibfnamefont {A.~I.}\ \bibnamefont {Goldman}}, \ and\ \bibinfo {author}
  {\bibfnamefont {D.~N.}\ \bibnamefont {Argyriou}},\ }\bibfield  {title}
  {\enquote {\bibinfo {title} {{Similarities between structural distortions
  under pressure and chemical doping in superconducting BaFe$_2$As$_2$}},}\
  }\href {\doibase 10.1038/nmat2443} {\bibfield  {journal} {\bibinfo  {journal}
  {Nature Mater.}\ }\textbf {\bibinfo {volume} {8}},\ \bibinfo {pages} {471}
  (\bibinfo {year} {2009})}\BibitemShut {NoStop}%
\bibitem [{\citenamefont {Kim}\ \emph {et~al.}(2011)\citenamefont {Kim},
  \citenamefont {Fernandes}, \citenamefont {Kreyssig}, \citenamefont {Kim},
  \citenamefont {Thaler}, \citenamefont {Bud'ko}, \citenamefont {Canfield},
  \citenamefont {McQueeney}, \citenamefont {Schmalian},\ and\ \citenamefont
  {Goldman}}]{Kim:2011b}%
  \BibitemOpen
  \bibfield  {author} {\bibinfo {author} {\bibfnamefont {M.~G.}\ \bibnamefont
  {Kim}}, \bibinfo {author} {\bibfnamefont {R.~M.}\ \bibnamefont {Fernandes}},
  \bibinfo {author} {\bibfnamefont {A.}~\bibnamefont {Kreyssig}}, \bibinfo
  {author} {\bibfnamefont {J.~W.}\ \bibnamefont {Kim}}, \bibinfo {author}
  {\bibfnamefont {A.}~\bibnamefont {Thaler}}, \bibinfo {author} {\bibfnamefont
  {S.~L.}\ \bibnamefont {Bud'ko}}, \bibinfo {author} {\bibfnamefont {P.~C.}\
  \bibnamefont {Canfield}}, \bibinfo {author} {\bibfnamefont {R.~J.}\
  \bibnamefont {McQueeney}}, \bibinfo {author} {\bibfnamefont {J.}~\bibnamefont
  {Schmalian}}, \ and\ \bibinfo {author} {\bibfnamefont {A.~I.}\ \bibnamefont
  {Goldman}},\ }\bibfield  {title} {\enquote {\bibinfo {title} {{Character of
  the structural and magnetic phase transitions in the parent and
  electron-doped BaFe${}_{2}$As${}_{2}$ compounds}},}\ }\href {\doibase
  10.1103/PhysRevB.83.134522} {\bibfield  {journal} {\bibinfo  {journal} {Phys.
  Rev. B}\ }\textbf {\bibinfo {volume} {83}},\ \bibinfo {pages} {134522}
  (\bibinfo {year} {2011})}\BibitemShut {NoStop}%
\bibitem [{\citenamefont {Rotter}\ \emph {et~al.}(2008)\citenamefont {Rotter},
  \citenamefont {Tegel},\ and\ \citenamefont {Johrendt}}]{Rotter:2008}%
  \BibitemOpen
  \bibfield  {author} {\bibinfo {author} {\bibfnamefont {Marianne}\
  \bibnamefont {Rotter}}, \bibinfo {author} {\bibfnamefont {Marcus}\
  \bibnamefont {Tegel}}, \ and\ \bibinfo {author} {\bibfnamefont {Dirk}\
  \bibnamefont {Johrendt}},\ }\bibfield  {title} {\enquote {\bibinfo {title}
  {{Superconductivity at 38\,K in the Iron Arsenide
  (Ba$_{1-x}$K$_x$)Fe$_2$As$_2$}},}\ }\href {\doibase
  10.1103/PhysRevLett.101.107006} {\bibfield  {journal} {\bibinfo  {journal}
  {Phys. Rev. Lett.}\ }\textbf {\bibinfo {volume} {101}},\ \bibinfo {eid}
  {107006} (\bibinfo {year} {2008})}\BibitemShut {NoStop}%
\bibitem [{\citenamefont {Chu}\ \emph {et~al.}(2010)\citenamefont {Chu},
  \citenamefont {Analytis}, \citenamefont {Greve}, \citenamefont {McMahon},
  \citenamefont {Islam}, \citenamefont {Yamamoto},\ and\ \citenamefont
  {Fisher}}]{Chu:2010}%
  \BibitemOpen
  \bibfield  {author} {\bibinfo {author} {\bibfnamefont {Jiun-Haw}\
  \bibnamefont {Chu}}, \bibinfo {author} {\bibfnamefont {James~G.}\
  \bibnamefont {Analytis}}, \bibinfo {author} {\bibfnamefont {Kristiaan~De}\
  \bibnamefont {Greve}}, \bibinfo {author} {\bibfnamefont {Peter~L.}\
  \bibnamefont {McMahon}}, \bibinfo {author} {\bibfnamefont {Zahirul}\
  \bibnamefont {Islam}}, \bibinfo {author} {\bibfnamefont {Yoshihisa}\
  \bibnamefont {Yamamoto}}, \ and\ \bibinfo {author} {\bibfnamefont {Ian~R.}\
  \bibnamefont {Fisher}},\ }\bibfield  {title} {\enquote {\bibinfo {title}
  {{In-Plane Resistivity Anisotropy in an Underdoped Iron Arsenide
  Superconductor}},}\ }\href {\doibase 10.1126/science.1190482} {\bibfield
  {journal} {\bibinfo  {journal} {Science}\ }\textbf {\bibinfo {volume}
  {329}},\ \bibinfo {pages} {824} (\bibinfo {year} {2010})}\BibitemShut
  {NoStop}%
\bibitem [{\citenamefont {Ying}\ \emph {et~al.}(2011)\citenamefont {Ying},
  \citenamefont {Wang}, \citenamefont {Wu}, \citenamefont {Xiang},
  \citenamefont {Liu}, \citenamefont {Yan}, \citenamefont {Wang}, \citenamefont
  {Zhang}, \citenamefont {Ye}, \citenamefont {Cheng}, \citenamefont {Hu},\ and\
  \citenamefont {Chen}}]{Ying:2011}%
  \BibitemOpen
  \bibfield  {author} {\bibinfo {author} {\bibfnamefont {J.~J.}\ \bibnamefont
  {Ying}}, \bibinfo {author} {\bibfnamefont {X.~F.}\ \bibnamefont {Wang}},
  \bibinfo {author} {\bibfnamefont {T.}~\bibnamefont {Wu}}, \bibinfo {author}
  {\bibfnamefont {Z.~J.}\ \bibnamefont {Xiang}}, \bibinfo {author}
  {\bibfnamefont {R.~H.}\ \bibnamefont {Liu}}, \bibinfo {author} {\bibfnamefont
  {Y.~J.}\ \bibnamefont {Yan}}, \bibinfo {author} {\bibfnamefont {A.~F.}\
  \bibnamefont {Wang}}, \bibinfo {author} {\bibfnamefont {M.}~\bibnamefont
  {Zhang}}, \bibinfo {author} {\bibfnamefont {G.~J.}\ \bibnamefont {Ye}},
  \bibinfo {author} {\bibfnamefont {P.}~\bibnamefont {Cheng}}, \bibinfo
  {author} {\bibfnamefont {J.~P.}\ \bibnamefont {Hu}}, \ and\ \bibinfo {author}
  {\bibfnamefont {X.~H.}\ \bibnamefont {Chen}},\ }\bibfield  {title} {\enquote
  {\bibinfo {title} {{Measurements of the Anisotropic In-Plane Resistivity of
  Underdoped FeAs-Based Pnictide Superconductors}},}\ }\href {\doibase
  10.1103/PhysRevLett.107.067001} {\bibfield  {journal} {\bibinfo  {journal}
  {Phys. Rev. Lett.}\ }\textbf {\bibinfo {volume} {107}},\ \bibinfo {pages}
  {067001} (\bibinfo {year} {2011})}\BibitemShut {NoStop}%
\bibitem [{\citenamefont {Dusza}\ \emph {et~al.}(2011)\citenamefont {Dusza},
  \citenamefont {Lucarelli}, \citenamefont {Pfuner}, \citenamefont {Chu},
  \citenamefont {Fisher},\ and\ \citenamefont {Degiorgi}}]{Dusza:2011}%
  \BibitemOpen
  \bibfield  {author} {\bibinfo {author} {\bibfnamefont {A.}~\bibnamefont
  {Dusza}}, \bibinfo {author} {\bibfnamefont {A.}~\bibnamefont {Lucarelli}},
  \bibinfo {author} {\bibfnamefont {F.}~\bibnamefont {Pfuner}}, \bibinfo
  {author} {\bibfnamefont {J.-H.}\ \bibnamefont {Chu}}, \bibinfo {author}
  {\bibfnamefont {I.~R.}\ \bibnamefont {Fisher}}, \ and\ \bibinfo {author}
  {\bibfnamefont {L.}~\bibnamefont {Degiorgi}},\ }\bibfield  {title} {\enquote
  {\bibinfo {title} {{Anisotropic charge dynamics in detwinned
  Ba(Fe$_{1-x}$Co$_x$)$_2$As$_2$}},}\ }\href {\doibase
  10.1209/0295-5075/93/37002} {\bibfield  {journal} {\bibinfo  {journal}
  {Europhys. Lett.}\ }\textbf {\bibinfo {volume} {93}},\ \bibinfo {pages}
  {37002} (\bibinfo {year} {2011})}\BibitemShut {NoStop}%
\bibitem [{\citenamefont {Dusza}\ \emph {et~al.}(2012)\citenamefont {Dusza},
  \citenamefont {Lucarelli}, \citenamefont {Sanna}, \citenamefont {Massidda},
  \citenamefont {Chu}, \citenamefont {Fisher},\ and\ \citenamefont
  {Degiorgi}}]{Dusza:2012}%
  \BibitemOpen
  \bibfield  {author} {\bibinfo {author} {\bibfnamefont {A.}~\bibnamefont
  {Dusza}}, \bibinfo {author} {\bibfnamefont {A.}~\bibnamefont {Lucarelli}},
  \bibinfo {author} {\bibfnamefont {A.}~\bibnamefont {Sanna}}, \bibinfo
  {author} {\bibfnamefont {S.}~\bibnamefont {Massidda}}, \bibinfo {author}
  {\bibfnamefont {J.-H.}\ \bibnamefont {Chu}}, \bibinfo {author} {\bibfnamefont
  {I.R.}\ \bibnamefont {Fisher}}, \ and\ \bibinfo {author} {\bibfnamefont
  {L.}~\bibnamefont {Degiorgi}},\ }\bibfield  {title} {\enquote {\bibinfo
  {title} {{Anisotropic in-plane optical conductivity in detwinned
  Ba(Fe$_{1-x}$Co$_x$)$_2$As$_2$}},}\ }\href {\doibase
  10.1088/1367-2630/14/2/023020} {\bibfield  {journal} {\bibinfo  {journal}
  {New J. Phys.}\ }\textbf {\bibinfo {volume} {14}},\ \bibinfo {pages} {023020}
  (\bibinfo {year} {2012})}\BibitemShut {NoStop}%
\bibitem [{\citenamefont {Nakajima}\ \emph {et~al.}(2011)\citenamefont
  {Nakajima}, \citenamefont {Liang}, \citenamefont {Ishida}, \citenamefont
  {Tomioka}, \citenamefont {Kihou}, \citenamefont {Lee}, \citenamefont {Iyo},
  \citenamefont {Eisaki}, \citenamefont {Kakeshita}, \citenamefont {Ito},\ and\
  \citenamefont {Uchida}}]{Nakajima:2011}%
  \BibitemOpen
  \bibfield  {author} {\bibinfo {author} {\bibfnamefont {M.}~\bibnamefont
  {Nakajima}}, \bibinfo {author} {\bibfnamefont {T.}~\bibnamefont {Liang}},
  \bibinfo {author} {\bibfnamefont {S.}~\bibnamefont {Ishida}}, \bibinfo
  {author} {\bibfnamefont {Y.}~\bibnamefont {Tomioka}}, \bibinfo {author}
  {\bibfnamefont {K.}~\bibnamefont {Kihou}}, \bibinfo {author} {\bibfnamefont
  {C.H.}\ \bibnamefont {Lee}}, \bibinfo {author} {\bibfnamefont
  {A.}~\bibnamefont {Iyo}}, \bibinfo {author} {\bibfnamefont {H.}~\bibnamefont
  {Eisaki}}, \bibinfo {author} {\bibfnamefont {T.}~\bibnamefont {Kakeshita}},
  \bibinfo {author} {\bibfnamefont {T.}~\bibnamefont {Ito}}, \ and\ \bibinfo
  {author} {\bibfnamefont {S.}~\bibnamefont {Uchida}},\ }\bibfield  {title}
  {\enquote {\bibinfo {title} {{Unprecedented anisotropic metallic state in
  undoped iron arsenide BaFe$_2$As$_2$ revealed by optical spectroscopy}},}\
  }\href {\doibase 10.1073/pnas.1100102108} {\bibfield  {journal} {\bibinfo
  {journal} {Proc. Natl. Acad. Sci.}\ }\textbf {\bibinfo {volume} {108}},\
  \bibinfo {pages} {12238} (\bibinfo {year} {2011})}\BibitemShut {NoStop}%
\bibitem [{\citenamefont {Yi}\ \emph {et~al.}(2011)\citenamefont {Yi},
  \citenamefont {Lu}, \citenamefont {Chu}, \citenamefont {Analytis},
  \citenamefont {Sorini}, \citenamefont {Kemper}, \citenamefont {Moritz},
  \citenamefont {Mo}, \citenamefont {Moore}, \citenamefont {Hashimoto},
  \citenamefont {Lee}, \citenamefont {Hussain}, \citenamefont {Devereaux},
  \citenamefont {Fisher},\ and\ \citenamefont {Shen}}]{Yi:2011}%
  \BibitemOpen
  \bibfield  {author} {\bibinfo {author} {\bibfnamefont {Ming}\ \bibnamefont
  {Yi}}, \bibinfo {author} {\bibfnamefont {Donghui}\ \bibnamefont {Lu}},
  \bibinfo {author} {\bibfnamefont {Jiun-Haw}\ \bibnamefont {Chu}}, \bibinfo
  {author} {\bibfnamefont {James~G.}\ \bibnamefont {Analytis}}, \bibinfo
  {author} {\bibfnamefont {Adam~P.}\ \bibnamefont {Sorini}}, \bibinfo {author}
  {\bibfnamefont {Alexander~F.}\ \bibnamefont {Kemper}}, \bibinfo {author}
  {\bibfnamefont {Brian}\ \bibnamefont {Moritz}}, \bibinfo {author}
  {\bibfnamefont {Sung-Kwan}\ \bibnamefont {Mo}}, \bibinfo {author}
  {\bibfnamefont {Rob~G.}\ \bibnamefont {Moore}}, \bibinfo {author}
  {\bibfnamefont {Makoto}\ \bibnamefont {Hashimoto}}, \bibinfo {author}
  {\bibfnamefont {Wei-Sheng}\ \bibnamefont {Lee}}, \bibinfo {author}
  {\bibfnamefont {Zahid}\ \bibnamefont {Hussain}}, \bibinfo {author}
  {\bibfnamefont {Thomas~P.}\ \bibnamefont {Devereaux}}, \bibinfo {author}
  {\bibfnamefont {Ian~R.}\ \bibnamefont {Fisher}}, \ and\ \bibinfo {author}
  {\bibfnamefont {Zhi-Xun}\ \bibnamefont {Shen}},\ }\bibfield  {title}
  {\enquote {\bibinfo {title} {{Symmetry-breaking orbital anisotropy observed
  for detwinned ${\rm Ba(Fe_{1-x}Co_x)_2As_2}$ above the spin density wave
  transition}},}\ }\href {\doibase 10.1073/pnas.1015572108} {\bibfield
  {journal} {\bibinfo  {journal} {Proc. Natl. Acad. Sci.}\ }\textbf {\bibinfo
  {volume} {108}},\ \bibinfo {pages} {6878--6883} (\bibinfo {year}
  {2011})}\BibitemShut {NoStop}%
\bibitem [{\citenamefont {Liang}\ \emph {et~al.}(2011)\citenamefont {Liang},
  \citenamefont {M.Nakajima}, \citenamefont {K.Kihou}, \citenamefont
  {Y.Tomioka}, \citenamefont {T.Ito}, \citenamefont {C.H.Lee}, \citenamefont
  {H.Kito}, \citenamefont {A.Iyo}, \citenamefont {H.Eisaki}, \citenamefont
  {Kakeshita},\ and\ \citenamefont {S.Uchida}}]{Liang:2011}%
  \BibitemOpen
  \bibfield  {author} {\bibinfo {author} {\bibfnamefont {T.}~\bibnamefont
  {Liang}}, \bibinfo {author} {\bibnamefont {M.Nakajima}}, \bibinfo {author}
  {\bibnamefont {K.Kihou}}, \bibinfo {author} {\bibnamefont {Y.Tomioka}},
  \bibinfo {author} {\bibnamefont {T.Ito}}, \bibinfo {author} {\bibnamefont
  {C.H.Lee}}, \bibinfo {author} {\bibnamefont {H.Kito}}, \bibinfo {author}
  {\bibnamefont {A.Iyo}}, \bibinfo {author} {\bibnamefont {H.Eisaki}}, \bibinfo
  {author} {\bibfnamefont {T.}~\bibnamefont {Kakeshita}}, \ and\ \bibinfo
  {author} {\bibnamefont {S.Uchida}},\ }\bibfield  {title} {\enquote {\bibinfo
  {title} {{Effects of uniaxial pressure and annealing on the resistivity of
  Ba(Fe$_{1-x}$Co$_x$)$_2$As$_2$}},}\ }\href {\doibase
  10.1016/j.jpcs.2010.10.080} {\bibfield  {journal} {\bibinfo  {journal} {J.
  Phys. Chem. Solids}\ }\textbf {\bibinfo {volume} {72}},\ \bibinfo {pages}
  {418} (\bibinfo {year} {2011})}\BibitemShut {NoStop}%
\bibitem [{\citenamefont {Blomberg}\ \emph {et~al.}(2012)\citenamefont
  {Blomberg}, \citenamefont {Kreyssig}, \citenamefont {Tanatar}, \citenamefont
  {Fernandes}, \citenamefont {Kim}, \citenamefont {Thaler}, \citenamefont
  {Schmalian}, \citenamefont {Bud'ko}, \citenamefont {Canfield}, \citenamefont
  {Goldman},\ and\ \citenamefont {Prozorov}}]{Blomberg:2012}%
  \BibitemOpen
  \bibfield  {author} {\bibinfo {author} {\bibfnamefont {E.~C.}\ \bibnamefont
  {Blomberg}}, \bibinfo {author} {\bibfnamefont {A.}~\bibnamefont {Kreyssig}},
  \bibinfo {author} {\bibfnamefont {M.~A.}\ \bibnamefont {Tanatar}}, \bibinfo
  {author} {\bibfnamefont {R.~M.}\ \bibnamefont {Fernandes}}, \bibinfo {author}
  {\bibfnamefont {M.~G.}\ \bibnamefont {Kim}}, \bibinfo {author} {\bibfnamefont
  {A.}~\bibnamefont {Thaler}}, \bibinfo {author} {\bibfnamefont
  {J.}~\bibnamefont {Schmalian}}, \bibinfo {author} {\bibfnamefont {S.~L.}\
  \bibnamefont {Bud'ko}}, \bibinfo {author} {\bibfnamefont {P.~C.}\
  \bibnamefont {Canfield}}, \bibinfo {author} {\bibfnamefont {A.~I.}\
  \bibnamefont {Goldman}}, \ and\ \bibinfo {author} {\bibfnamefont
  {R.}~\bibnamefont {Prozorov}},\ }\bibfield  {title} {\enquote {\bibinfo
  {title} {{Effect of tensile stress on the in-plane resistivity anisotropy in
  BaFe$_{2}$As$_{2}$}},}\ }\href {\doibase 10.1103/PhysRevB.85.144509}
  {\bibfield  {journal} {\bibinfo  {journal} {Phys. Rev. B}\ }\textbf {\bibinfo
  {volume} {85}},\ \bibinfo {pages} {144509} (\bibinfo {year}
  {2012})}\BibitemShut {NoStop}%
\bibitem [{\citenamefont {Perdew}\ \emph {et~al.}(1996)\citenamefont {Perdew},
  \citenamefont {Burke},\ and\ \citenamefont {Ernzerhof}}]{Perdew:1996}%
  \BibitemOpen
  \bibfield  {author} {\bibinfo {author} {\bibfnamefont {John~P.}\ \bibnamefont
  {Perdew}}, \bibinfo {author} {\bibfnamefont {Kieron}\ \bibnamefont {Burke}},
  \ and\ \bibinfo {author} {\bibfnamefont {Matthias}\ \bibnamefont
  {Ernzerhof}},\ }\bibfield  {title} {\enquote {\bibinfo {title} {{Generalized
  Gradient Approximation Made Simple}},}\ }\href {\doibase
  10.1103/PhysRevLett.77.3865} {\bibfield  {journal} {\bibinfo  {journal}
  {Phys. Rev. Lett.}\ }\textbf {\bibinfo {volume} {77}},\ \bibinfo {pages}
  {3865--3868} (\bibinfo {year} {1996})}\BibitemShut {NoStop}%
\bibitem [{\citenamefont {Togo}\ \emph {et~al.}(2008)\citenamefont {Togo},
  \citenamefont {Oba},\ and\ \citenamefont {Tanaka}}]{Togo:2008}%
  \BibitemOpen
  \bibfield  {author} {\bibinfo {author} {\bibfnamefont {Atsushi}\ \bibnamefont
  {Togo}}, \bibinfo {author} {\bibfnamefont {Fumiyasu}\ \bibnamefont {Oba}}, \
  and\ \bibinfo {author} {\bibfnamefont {Isao}\ \bibnamefont {Tanaka}},\
  }\bibfield  {title} {\enquote {\bibinfo {title} {{First-principles
  calculations of the ferroelastic transition between rutile-type and
  ${\text{CaCl}}_{2}$-type ${\text{SiO}}_{2}$ at high pressures}},}\ }\href
  {\doibase 10.1103/PhysRevB.78.134106} {\bibfield  {journal} {\bibinfo
  {journal} {Phys. Rev. B}\ }\textbf {\bibinfo {volume} {78}},\ \bibinfo
  {pages} {134106} (\bibinfo {year} {2008})}\BibitemShut {NoStop}%
\bibitem [{\citenamefont {Togo}\ and\ \citenamefont
  {Tanaka}(2015)}]{Togo:2015}%
  \BibitemOpen
  \bibfield  {author} {\bibinfo {author} {\bibfnamefont {Atsushi}\ \bibnamefont
  {Togo}}\ and\ \bibinfo {author} {\bibfnamefont {Isao}\ \bibnamefont
  {Tanaka}},\ }\bibfield  {title} {\enquote {\bibinfo {title} {{First
  principles phonon calculations in materials science}},}\ }\href {\doibase
  10.1016/j.scriptamat.2015.07.021} {\bibfield  {journal} {\bibinfo  {journal}
  {Scr. Mater.}\ }\textbf {\bibinfo {volume} {108}},\ \bibinfo {pages} {1--5}
  (\bibinfo {year} {2015})}\BibitemShut {NoStop}%
\bibitem [{\citenamefont {Parlinski}\ \emph {et~al.}(1997)\citenamefont
  {Parlinski}, \citenamefont {Li},\ and\ \citenamefont
  {Kawazoe}}]{Parlinski:1997}%
  \BibitemOpen
  \bibfield  {author} {\bibinfo {author} {\bibfnamefont {K.}~\bibnamefont
  {Parlinski}}, \bibinfo {author} {\bibfnamefont {Z.~Q.}\ \bibnamefont {Li}}, \
  and\ \bibinfo {author} {\bibfnamefont {Y.}~\bibnamefont {Kawazoe}},\
  }\bibfield  {title} {\enquote {\bibinfo {title} {{First-Principles
  Determination of the Soft Mode in Cubic ${\mathrm{ZrO}}_{2}$}},}\ }\href
  {\doibase 10.1103/PhysRevLett.78.4063} {\bibfield  {journal} {\bibinfo
  {journal} {Phys. Rev. Lett.}\ }\textbf {\bibinfo {volume} {78}},\ \bibinfo
  {pages} {4063--4066} (\bibinfo {year} {1997})}\BibitemShut {NoStop}%
\bibitem [{\citenamefont {Bl\"ochl}(1994)}]{Bloechl:1994}%
  \BibitemOpen
  \bibfield  {author} {\bibinfo {author} {\bibfnamefont {P.~E.}\ \bibnamefont
  {Bl\"ochl}},\ }\bibfield  {title} {\enquote {\bibinfo {title} {{Projector
  augmented-wave method}},}\ }\href {\doibase 10.1103/PhysRevB.50.17953}
  {\bibfield  {journal} {\bibinfo  {journal} {Phys. Rev. B}\ }\textbf {\bibinfo
  {volume} {50}},\ \bibinfo {pages} {17953--17979} (\bibinfo {year}
  {1994})}\BibitemShut {NoStop}%
\bibitem [{\citenamefont {Kresse}\ and\ \citenamefont
  {Hafner}(1993)}]{Kresse:1993}%
  \BibitemOpen
  \bibfield  {author} {\bibinfo {author} {\bibfnamefont {G.}~\bibnamefont
  {Kresse}}\ and\ \bibinfo {author} {\bibfnamefont {J.}~\bibnamefont
  {Hafner}},\ }\bibfield  {title} {\enquote {\bibinfo {title} {{Ab initio
  molecular dynamics for liquid metals}},}\ }\href {\doibase
  10.1103/PhysRevB.47.558} {\bibfield  {journal} {\bibinfo  {journal} {Phys.
  Rev. B}\ }\textbf {\bibinfo {volume} {47}},\ \bibinfo {pages} {558--561}
  (\bibinfo {year} {1993})}\BibitemShut {NoStop}%
\bibitem [{\citenamefont {{Kresse, G. and Furthm\"uller,
  J.}}(1996)}]{Kresse:1996}%
  \BibitemOpen
  \bibfield  {author} {\bibinfo {author} {\bibnamefont {{Kresse, G. and
  Furthm\"uller, J.}}},\ }\bibfield  {title} {\enquote {\bibinfo {title}
  {{Efficient iterative schemes for ab initio total-energy calculations using a
  plane-wave basis set}},}\ }\href {\doibase 10.1103/PhysRevB.54.11169}
  {\bibfield  {journal} {\bibinfo  {journal} {Phys. Rev. B}\ }\textbf {\bibinfo
  {volume} {54}},\ \bibinfo {pages} {11169--11186} (\bibinfo {year}
  {1996})}\BibitemShut {NoStop}%
\bibitem [{\citenamefont {Kresse}\ and\ \citenamefont
  {Furthm\"{u}ller}(1996)}]{Kresse:1996b}%
  \BibitemOpen
  \bibfield  {author} {\bibinfo {author} {\bibfnamefont {G.}~\bibnamefont
  {Kresse}}\ and\ \bibinfo {author} {\bibfnamefont {J.}~\bibnamefont
  {Furthm\"{u}ller}},\ }\bibfield  {title} {\enquote {\bibinfo {title}
  {{Efficiency of ab-initio total energy calculations for metals and
  semiconductors using a plane-wave basis set}},}\ }\href {\doibase
  http://dx.doi.org/10.1016/0927-0256(96)00008-0} {\bibfield  {journal}
  {\bibinfo  {journal} {Comput. Mater. Sci.}\ }\textbf {\bibinfo {volume}
  {6}},\ \bibinfo {pages} {15--50} (\bibinfo {year} {1996})}\BibitemShut
  {NoStop}%
\bibitem [{Note1()}]{Note1}%
  \BibitemOpen
  \bibinfo {note} {The local correlations in the tetragonal phase are of the
  stripe type; however, we had to use a pattern that does not break the
  symmetry, and it is known \cite {Mazin:2008a} that the difference in the
  calculated elastic properties calculated within different magnetic orders is
  much smaller than between magnetic and nonmagnetic calculations}\BibitemShut
  {NoStop}%
\bibitem [{\citenamefont {Ambrosch-Draxl}\ and\ \citenamefont
  {Sofo}(2006)}]{Ambrosch-Draxl:2006}%
  \BibitemOpen
  \bibfield  {author} {\bibinfo {author} {\bibfnamefont {Claudia}\ \bibnamefont
  {Ambrosch-Draxl}}\ and\ \bibinfo {author} {\bibfnamefont {Jorge~O.}\
  \bibnamefont {Sofo}},\ }\bibfield  {title} {\enquote {\bibinfo {title}
  {{Linear optical properties of solids within the full-potential linearized
  augmented planewave method}},}\ }\href {\doibase 10.1016/j.cpc.2006.03.005}
  {\bibfield  {journal} {\bibinfo  {journal} {Comput. Phys. Commun.}\ }\textbf
  {\bibinfo {volume} {175}},\ \bibinfo {pages} {1--14} (\bibinfo {year}
  {2006})}\BibitemShut {NoStop}%
\bibitem [{\citenamefont {Blaha}\ \emph {et~al.}(2001)\citenamefont {Blaha},
  \citenamefont {Schwarz}, \citenamefont {Madsen}, \citenamefont {Kvasnicka},\
  and\ \citenamefont {Luitz}}]{Blaha:2001}%
  \BibitemOpen
  \bibfield  {author} {\bibinfo {author} {\bibfnamefont {Peter}\ \bibnamefont
  {Blaha}}, \bibinfo {author} {\bibfnamefont {Karlheinz}\ \bibnamefont
  {Schwarz}}, \bibinfo {author} {\bibfnamefont {G.~K.~H.}\ \bibnamefont
  {Madsen}}, \bibinfo {author} {\bibfnamefont {D.}~\bibnamefont {Kvasnicka}}, \
  and\ \bibinfo {author} {\bibfnamefont {J.}~\bibnamefont {Luitz}},\ }\href
  {http://susi.theochem.tuwien.ac.at/} {\enquote {\bibinfo {title} {{An
  Augmented Plane Wave Plus Local Orbitals Program for Calculating Crystal
  Properties}},}\ } (\bibinfo {year} {2001})\BibitemShut {NoStop}%
\bibitem [{\citenamefont {Ren}\ \emph {et~al.}(2015)\citenamefont {Ren},
  \citenamefont {Duan}, \citenamefont {Hu}, \citenamefont {Li}, \citenamefont
  {Zhang}, \citenamefont {Luo}, \citenamefont {Dai},\ and\ \citenamefont
  {Li}}]{RenX:2015}%
  \BibitemOpen
  \bibfield  {author} {\bibinfo {author} {\bibfnamefont {Xiao}\ \bibnamefont
  {Ren}}, \bibinfo {author} {\bibfnamefont {Lian}\ \bibnamefont {Duan}},
  \bibinfo {author} {\bibfnamefont {Yuwen}\ \bibnamefont {Hu}}, \bibinfo
  {author} {\bibfnamefont {Jiarui}\ \bibnamefont {Li}}, \bibinfo {author}
  {\bibfnamefont {Rui}\ \bibnamefont {Zhang}}, \bibinfo {author} {\bibfnamefont
  {Huiqian}\ \bibnamefont {Luo}}, \bibinfo {author} {\bibfnamefont {Pengcheng}\
  \bibnamefont {Dai}}, \ and\ \bibinfo {author} {\bibfnamefont {Yuan}\
  \bibnamefont {Li}},\ }\bibfield  {title} {\enquote {\bibinfo {title}
  {{Nematic Crossover in ${\mathrm{BaFe}}_{2}{\mathrm{As}}_{2}$ under Uniaxial
  Stress}},}\ }\href {\doibase 10.1103/PhysRevLett.115.197002} {\bibfield
  {journal} {\bibinfo  {journal} {Phys. Rev. Lett.}\ }\textbf {\bibinfo
  {volume} {115}},\ \bibinfo {pages} {197002} (\bibinfo {year}
  {2015})}\BibitemShut {NoStop}%
\bibitem [{\citenamefont {Zhang}\ \emph {et~al.}(2016)\citenamefont {Zhang},
  \citenamefont {Sefat}, \citenamefont {Ding}, \citenamefont {Richard},\ and\
  \citenamefont {Blumberg}}]{ZhangWL:2016}%
  \BibitemOpen
  \bibfield  {author} {\bibinfo {author} {\bibfnamefont {W.-L.}\ \bibnamefont
  {Zhang}}, \bibinfo {author} {\bibfnamefont {Athena~S.}\ \bibnamefont
  {Sefat}}, \bibinfo {author} {\bibfnamefont {H.}~\bibnamefont {Ding}},
  \bibinfo {author} {\bibfnamefont {P.}~\bibnamefont {Richard}}, \ and\
  \bibinfo {author} {\bibfnamefont {G.}~\bibnamefont {Blumberg}},\ }\bibfield
  {title} {\enquote {\bibinfo {title} {{Stress-induced nematicity in
  ${\mathrm{EuFe}}_{2}{\mathrm{As}}_{2}$ studied by Raman spectroscopy}},}\
  }\href {\doibase 10.1103/PhysRevB.94.014513} {\bibfield  {journal} {\bibinfo
  {journal} {Phys. Rev. B}\ }\textbf {\bibinfo {volume} {94}},\ \bibinfo
  {pages} {014513} (\bibinfo {year} {2016})}\BibitemShut {NoStop}%
\bibitem [{\citenamefont {Choi}\ \emph {et~al.}(2008)\citenamefont {Choi},
  \citenamefont {Wulferding}, \citenamefont {Lemmens}, \citenamefont {Ni},
  \citenamefont {Bud'ko},\ and\ \citenamefont {Canfield}}]{Choi:2008}%
  \BibitemOpen
  \bibfield  {author} {\bibinfo {author} {\bibfnamefont {K.-Y.}\ \bibnamefont
  {Choi}}, \bibinfo {author} {\bibfnamefont {D.}~\bibnamefont {Wulferding}},
  \bibinfo {author} {\bibfnamefont {P.}~\bibnamefont {Lemmens}}, \bibinfo
  {author} {\bibfnamefont {N.}~\bibnamefont {Ni}}, \bibinfo {author}
  {\bibfnamefont {S.~L.}\ \bibnamefont {Bud'ko}}, \ and\ \bibinfo {author}
  {\bibfnamefont {P.~C.}\ \bibnamefont {Canfield}},\ }\bibfield  {title}
  {\enquote {\bibinfo {title} {{Lattice and electronic anomalies of
  CaFe$_2$As$_2$ studied by Raman spectroscopy}},}\ }\href {\doibase
  10.1103/PhysRevB.78.212503} {\bibfield  {journal} {\bibinfo  {journal} {Phys.
  Rev. B}\ }\textbf {\bibinfo {volume} {78}},\ \bibinfo {pages} {212503}
  (\bibinfo {year} {2008})}\BibitemShut {NoStop}%
\bibitem [{\citenamefont {Chauvi\`ere}\ \emph {et~al.}(2010)\citenamefont
  {Chauvi\`ere}, \citenamefont {Gallais}, \citenamefont {Cazayous},
  \citenamefont {M\'easson}, \citenamefont {Sacuto}, \citenamefont {Colson},\
  and\ \citenamefont {Forget}}]{Chauviere:2010}%
  \BibitemOpen
  \bibfield  {author} {\bibinfo {author} {\bibfnamefont {L.}~\bibnamefont
  {Chauvi\`ere}}, \bibinfo {author} {\bibfnamefont {Y.}~\bibnamefont
  {Gallais}}, \bibinfo {author} {\bibfnamefont {M.}~\bibnamefont {Cazayous}},
  \bibinfo {author} {\bibfnamefont {M.~A.}\ \bibnamefont {M\'easson}}, \bibinfo
  {author} {\bibfnamefont {A.}~\bibnamefont {Sacuto}}, \bibinfo {author}
  {\bibfnamefont {D.}~\bibnamefont {Colson}}, \ and\ \bibinfo {author}
  {\bibfnamefont {A.}~\bibnamefont {Forget}},\ }\bibfield  {title} {\enquote
  {\bibinfo {title} {{Impact of the spin-density-wave order on the
  superconducting gap of Ba(Fe$_{1-x}$Co$_{x}$)$_{2}$As$_{2}$}},}\ }\href
  {\doibase 10.1103/PhysRevB.82.180521} {\bibfield  {journal} {\bibinfo
  {journal} {Phys. Rev. B}\ }\textbf {\bibinfo {volume} {82}},\ \bibinfo
  {pages} {180521} (\bibinfo {year} {2010})}\BibitemShut {NoStop}%
\bibitem [{\citenamefont {Sugai}\ \emph {et~al.}(2012)\citenamefont {Sugai},
  \citenamefont {Mizuno}, \citenamefont {Watanabe}, \citenamefont {Kawaguchi},
  \citenamefont {Takenaka}, \citenamefont {Ikuta}, \citenamefont {Takayanagi},
  \citenamefont {Hayamizu},\ and\ \citenamefont {Sone}}]{Sugai:2012}%
  \BibitemOpen
  \bibfield  {author} {\bibinfo {author} {\bibfnamefont {Shunji}\ \bibnamefont
  {Sugai}}, \bibinfo {author} {\bibfnamefont {Yuki}\ \bibnamefont {Mizuno}},
  \bibinfo {author} {\bibfnamefont {Ryoutarou}\ \bibnamefont {Watanabe}},
  \bibinfo {author} {\bibfnamefont {Takahiko}\ \bibnamefont {Kawaguchi}},
  \bibinfo {author} {\bibfnamefont {Koshi}\ \bibnamefont {Takenaka}}, \bibinfo
  {author} {\bibfnamefont {Hiroshi}\ \bibnamefont {Ikuta}}, \bibinfo {author}
  {\bibfnamefont {Yasumasa}\ \bibnamefont {Takayanagi}}, \bibinfo {author}
  {\bibfnamefont {Naoki}\ \bibnamefont {Hayamizu}}, \ and\ \bibinfo {author}
  {\bibfnamefont {Yasuhiro}\ \bibnamefont {Sone}},\ }\bibfield  {title}
  {\enquote {\bibinfo {title} {{Spin-Density-Wave Gap with Dirac Nodes and
  Two-Magnon Raman Scattering in BaFe$_2$As$_2$}},}\ }\href {\doibase
  10.1143/JPSJ.81.024718} {\bibfield  {journal} {\bibinfo  {journal} {J. Phys.
  Soc. Japan}\ }\textbf {\bibinfo {volume} {81}},\ \bibinfo {pages} {024718}
  (\bibinfo {year} {2012})}\BibitemShut {NoStop}%
\bibitem [{\citenamefont {{Kretzschmar}}\ \emph {et~al.}(2016)\citenamefont
  {{Kretzschmar}}, \citenamefont {{B{\"o}hm}}, \citenamefont
  {{Karahasanovi{\'c}}}, \citenamefont {{Muschler}}, \citenamefont {{Baum}},
  \citenamefont {{Jost}}, \citenamefont {{Schmalian}}, \citenamefont
  {{Caprara}}, \citenamefont {{Grilli}}, \citenamefont {{Di Castro}},
  \citenamefont {{Analytis}}, \citenamefont {{Chu}}, \citenamefont {{Fisher}},\
  and\ \citenamefont {{Hackl}}}]{Kretzschmar:2016}%
  \BibitemOpen
  \bibfield  {author} {\bibinfo {author} {\bibfnamefont {F.}~\bibnamefont
  {{Kretzschmar}}}, \bibinfo {author} {\bibfnamefont {T.}~\bibnamefont
  {{B{\"o}hm}}}, \bibinfo {author} {\bibfnamefont {U.}~\bibnamefont
  {{Karahasanovi{\'c}}}}, \bibinfo {author} {\bibfnamefont {B.}~\bibnamefont
  {{Muschler}}}, \bibinfo {author} {\bibfnamefont {A.}~\bibnamefont {{Baum}}},
  \bibinfo {author} {\bibfnamefont {D.}~\bibnamefont {{Jost}}}, \bibinfo
  {author} {\bibfnamefont {J.}~\bibnamefont {{Schmalian}}}, \bibinfo {author}
  {\bibfnamefont {S.}~\bibnamefont {{Caprara}}}, \bibinfo {author}
  {\bibfnamefont {M.}~\bibnamefont {{Grilli}}}, \bibinfo {author}
  {\bibfnamefont {C.}~\bibnamefont {{Di Castro}}}, \bibinfo {author}
  {\bibfnamefont {J.~H.}\ \bibnamefont {{Analytis}}}, \bibinfo {author}
  {\bibfnamefont {J.-H.}\ \bibnamefont {{Chu}}}, \bibinfo {author}
  {\bibfnamefont {I.~R.}\ \bibnamefont {{Fisher}}}, \ and\ \bibinfo {author}
  {\bibfnamefont {R.}~\bibnamefont {{Hackl}}},\ }\bibfield  {title} {\enquote
  {\bibinfo {title} {{Critical spin fluctuations and the origin of nematic
  order in ${\rm Ba(Fe_{1-x}Co_x)_2As_2}$}},}\ }\href {\doibase
  10.1038/NPHYS3634} {\bibfield  {journal} {\bibinfo  {journal} {Nat. Phys.}\
  }\textbf {\bibinfo {volume} {12}},\ \bibinfo {pages} {560--563} (\bibinfo
  {year} {2016})}\BibitemShut {NoStop}%
\bibitem [{\citenamefont {Thorsm\o{}lle}\ \emph {et~al.}(2016)\citenamefont
  {Thorsm\o{}lle}, \citenamefont {Khodas}, \citenamefont {Yin}, \citenamefont
  {Zhang}, \citenamefont {Carr}, \citenamefont {Dai},\ and\ \citenamefont
  {Blumberg}}]{Thorsmolle:2016}%
  \BibitemOpen
  \bibfield  {author} {\bibinfo {author} {\bibfnamefont {V.~K.}\ \bibnamefont
  {Thorsm\o{}lle}}, \bibinfo {author} {\bibfnamefont {M.}~\bibnamefont
  {Khodas}}, \bibinfo {author} {\bibfnamefont {Z.~P.}\ \bibnamefont {Yin}},
  \bibinfo {author} {\bibfnamefont {Chenglin}\ \bibnamefont {Zhang}}, \bibinfo
  {author} {\bibfnamefont {S.~V.}\ \bibnamefont {Carr}}, \bibinfo {author}
  {\bibfnamefont {Pengcheng}\ \bibnamefont {Dai}}, \ and\ \bibinfo {author}
  {\bibfnamefont {G.}~\bibnamefont {Blumberg}},\ }\bibfield  {title} {\enquote
  {\bibinfo {title} {Critical quadrupole fluctuations and collective modes in
  iron pnictide superconductors},}\ }\href {\doibase
  10.1103/PhysRevB.93.054515} {\bibfield  {journal} {\bibinfo  {journal} {Phys.
  Rev. B}\ }\textbf {\bibinfo {volume} {93}},\ \bibinfo {pages} {054515}
  (\bibinfo {year} {2016})}\BibitemShut {NoStop}%
\bibitem [{\citenamefont {Cardona}(1982)}]{Cardona1982_ResonancePhenomena}%
  \BibitemOpen
  \bibfield  {author} {\bibinfo {author} {\bibfnamefont {M.}~\bibnamefont
  {Cardona}},\ }\enquote {\bibinfo {title} {{Resonance Phenomena}},}\ \
  (\bibinfo  {publisher} {Springer-Verlag Berlin Heidelberg},\ \bibinfo {year}
  {1982})\ Chap.~\bibinfo {chapter} {2}, pp.\ \bibinfo {pages} {19--178},\
  \bibinfo {edition} {1st}\ ed.\BibitemShut {Stop}%
\bibitem [{\citenamefont {Baek}\ \emph {et~al.}(2014)\citenamefont {Baek},
  \citenamefont {Efremov}, \citenamefont {Ok}, \citenamefont {Kim},
  \citenamefont {van~den Brink},\ and\ \citenamefont {B\"uchner}}]{Baek:2014}%
  \BibitemOpen
  \bibfield  {author} {\bibinfo {author} {\bibfnamefont {S.-H.}\ \bibnamefont
  {Baek}}, \bibinfo {author} {\bibfnamefont {D.~V.}\ \bibnamefont {Efremov}},
  \bibinfo {author} {\bibfnamefont {J.~M.}\ \bibnamefont {Ok}}, \bibinfo
  {author} {\bibfnamefont {J.~S.}\ \bibnamefont {Kim}}, \bibinfo {author}
  {\bibfnamefont {Jeroen}\ \bibnamefont {van~den Brink}}, \ and\ \bibinfo
  {author} {\bibfnamefont {B.}~\bibnamefont {B\"uchner}},\ }\bibfield  {title}
  {\enquote {\bibinfo {title} {{Orbital-driven nematicity in FeSe}},}\ }\href
  {\doibase 10.1038/nmat4138} {\bibfield  {journal} {\bibinfo  {journal} {Nat.
  Mater.}\ }\textbf {\bibinfo {volume} {14}},\ \bibinfo {pages} {210--214}
  (\bibinfo {year} {2014})}\BibitemShut {NoStop}%
\bibitem [{\citenamefont {Muschler}\ \emph {et~al.}(2010)\citenamefont
  {Muschler}, \citenamefont {Prestel}, \citenamefont {Tassini}, \citenamefont
  {Hackl}, \citenamefont {Lambacher}, \citenamefont {Erb}, \citenamefont
  {Komiya}, \citenamefont {Ando}, \citenamefont {Peets}, \citenamefont {Hardy},
  \citenamefont {Liang},\ and\ \citenamefont {Bonn}}]{Muschler:2010a}%
  \BibitemOpen
  \bibfield  {author} {\bibinfo {author} {\bibfnamefont {B.}~\bibnamefont
  {Muschler}}, \bibinfo {author} {\bibfnamefont {W.}~\bibnamefont {Prestel}},
  \bibinfo {author} {\bibfnamefont {L.}~\bibnamefont {Tassini}}, \bibinfo
  {author} {\bibfnamefont {R.}~\bibnamefont {Hackl}}, \bibinfo {author}
  {\bibfnamefont {M.}~\bibnamefont {Lambacher}}, \bibinfo {author}
  {\bibfnamefont {A.}~\bibnamefont {Erb}}, \bibinfo {author} {\bibfnamefont
  {Seiki}\ \bibnamefont {Komiya}}, \bibinfo {author} {\bibfnamefont {Yoichi}\
  \bibnamefont {Ando}}, \bibinfo {author} {\bibfnamefont {D.C.}\ \bibnamefont
  {Peets}}, \bibinfo {author} {\bibfnamefont {W.N.}\ \bibnamefont {Hardy}},
  \bibinfo {author} {\bibfnamefont {R.}~\bibnamefont {Liang}}, \ and\ \bibinfo
  {author} {\bibfnamefont {D.A.}\ \bibnamefont {Bonn}},\ }\bibfield  {title}
  {\enquote {\bibinfo {title} {{Electron interactions and charge ordering in
  CuO$_2$ compounds}},}\ }\href {\doibase 10.1140/epjst/e2010-01302-4}
  {\bibfield  {journal} {\bibinfo  {journal} {Eur. Phys. J. Special Topics}\
  }\textbf {\bibinfo {volume} {188}},\ \bibinfo {pages} {131} (\bibinfo {year}
  {2010})}\BibitemShut {NoStop}%
\bibitem [{\citenamefont {Strach}\ \emph {et~al.}(1998)\citenamefont {Strach},
  \citenamefont {Brunen}, \citenamefont {Lederle}, \citenamefont {Zegenhagen},\
  and\ \citenamefont {Cardona}}]{Strach:1998}%
  \BibitemOpen
  \bibfield  {author} {\bibinfo {author} {\bibfnamefont {T.}~\bibnamefont
  {Strach}}, \bibinfo {author} {\bibfnamefont {J.}~\bibnamefont {Brunen}},
  \bibinfo {author} {\bibfnamefont {B.}~\bibnamefont {Lederle}}, \bibinfo
  {author} {\bibfnamefont {J.}~\bibnamefont {Zegenhagen}}, \ and\ \bibinfo
  {author} {\bibfnamefont {M.}~\bibnamefont {Cardona}},\ }\bibfield  {title}
  {\enquote {\bibinfo {title} {{Determination of the phase difference between
  the Raman tensor elements of the ${A}_{1g}$-like phonons in
  ${\mathrm{SmBa}}_{2}{\mathrm{Cu}}_{3}{\mathrm{O}}_{7-\ensuremath{\delta}}$}},}\
  }\href {\doibase 10.1103/PhysRevB.57.1292} {\bibfield  {journal} {\bibinfo
  {journal} {Phys. Rev. B}\ }\textbf {\bibinfo {volume} {57}},\ \bibinfo
  {pages} {1292--1297} (\bibinfo {year} {1998})}\BibitemShut {NoStop}%
\bibitem [{\citenamefont {Ambrosch-Draxl}\ \emph {et~al.}(2002)\citenamefont
  {Ambrosch-Draxl}, \citenamefont {Auer}, \citenamefont {Kouba}, \citenamefont
  {Sherman}, \citenamefont {Knoll},\ and\ \citenamefont
  {Mayer}}]{Ambrosch:2002}%
  \BibitemOpen
  \bibfield  {author} {\bibinfo {author} {\bibfnamefont {C.}~\bibnamefont
  {Ambrosch-Draxl}}, \bibinfo {author} {\bibfnamefont {H.}~\bibnamefont
  {Auer}}, \bibinfo {author} {\bibfnamefont {R.}~\bibnamefont {Kouba}},
  \bibinfo {author} {\bibfnamefont {E.~Ya.}\ \bibnamefont {Sherman}}, \bibinfo
  {author} {\bibfnamefont {P.}~\bibnamefont {Knoll}}, \ and\ \bibinfo {author}
  {\bibfnamefont {M.}~\bibnamefont {Mayer}},\ }\bibfield  {title} {\enquote
  {\bibinfo {title} {{Raman scattering in
  ${\mathrm{YBa}}_{2}{\mathrm{Cu}}_{3}{\mathrm{O}}_{7}:$ A comprehensive
  theoretical study in comparison with experiments}},}\ }\href {\doibase
  10.1103/PhysRevB.65.064501} {\bibfield  {journal} {\bibinfo  {journal} {Phys.
  Rev. B}\ }\textbf {\bibinfo {volume} {65}},\ \bibinfo {pages} {064501}
  (\bibinfo {year} {2002})}\BibitemShut {NoStop}%
\bibitem [{\citenamefont {Skornyakov}\ \emph {et~al.}(2009)\citenamefont
  {Skornyakov}, \citenamefont {Efremov}, \citenamefont {Skorikov},
  \citenamefont {Korotin}, \citenamefont {Izyumov}, \citenamefont {Anisimov},
  \citenamefont {Kozhevnikov},\ and\ \citenamefont
  {Vollhardt}}]{Skornyakov:2009}%
  \BibitemOpen
  \bibfield  {author} {\bibinfo {author} {\bibfnamefont {S.~L.}\ \bibnamefont
  {Skornyakov}}, \bibinfo {author} {\bibfnamefont {A.~V.}\ \bibnamefont
  {Efremov}}, \bibinfo {author} {\bibfnamefont {N.~A.}\ \bibnamefont
  {Skorikov}}, \bibinfo {author} {\bibfnamefont {M.~A.}\ \bibnamefont
  {Korotin}}, \bibinfo {author} {\bibfnamefont {Yu.~A.}\ \bibnamefont
  {Izyumov}}, \bibinfo {author} {\bibfnamefont {V.~I.}\ \bibnamefont
  {Anisimov}}, \bibinfo {author} {\bibfnamefont {A.~V.}\ \bibnamefont
  {Kozhevnikov}}, \ and\ \bibinfo {author} {\bibfnamefont {D.}~\bibnamefont
  {Vollhardt}},\ }\bibfield  {title} {\enquote {\bibinfo {title}
  {{Classification of the electronic correlation strength in the iron
  pnictides: The case of the parent compound
  ${\text{BaFe}}_{2}{\text{As}}_{2}$}},}\ }\href {\doibase
  10.1103/PhysRevB.80.092501} {\bibfield  {journal} {\bibinfo  {journal} {Phys.
  Rev. B}\ }\textbf {\bibinfo {volume} {80}},\ \bibinfo {pages} {092501}
  (\bibinfo {year} {2009})}\BibitemShut {NoStop}%
\bibitem [{\citenamefont {Yao}\ \emph {et~al.}(2011)\citenamefont {Yao},
  \citenamefont {Schmalian}, \citenamefont {Wang}, \citenamefont {Ho},\ and\
  \citenamefont {Kotliar}}]{Yao:2011}%
  \BibitemOpen
  \bibfield  {author} {\bibinfo {author} {\bibfnamefont {Y.~X.}\ \bibnamefont
  {Yao}}, \bibinfo {author} {\bibfnamefont {J.}~\bibnamefont {Schmalian}},
  \bibinfo {author} {\bibfnamefont {C.~Z.}\ \bibnamefont {Wang}}, \bibinfo
  {author} {\bibfnamefont {K.~M.}\ \bibnamefont {Ho}}, \ and\ \bibinfo {author}
  {\bibfnamefont {G.}~\bibnamefont {Kotliar}},\ }\bibfield  {title} {\enquote
  {\bibinfo {title} {{Comparative study of the electronic and magnetic
  properties of BaFe$_{2}$As$_{2}$ and BaMn$_{2}$As$_{2}$ using the Gutzwiller
  approximation}},}\ }\href {\doibase 10.1103/PhysRevB.84.245112} {\bibfield
  {journal} {\bibinfo  {journal} {Phys. Rev. B}\ }\textbf {\bibinfo {volume}
  {84}},\ \bibinfo {pages} {245112} (\bibinfo {year} {2011})}\BibitemShut
  {NoStop}%
\bibitem [{\citenamefont {Ferber}\ \emph {et~al.}(2012)\citenamefont {Ferber},
  \citenamefont {Foyevtsova}, \citenamefont {Valent\'{\i}},\ and\ \citenamefont
  {Jeschke}}]{Ferber:2012}%
  \BibitemOpen
  \bibfield  {author} {\bibinfo {author} {\bibfnamefont {Johannes}\
  \bibnamefont {Ferber}}, \bibinfo {author} {\bibfnamefont {Kateryna}\
  \bibnamefont {Foyevtsova}}, \bibinfo {author} {\bibfnamefont {Roser}\
  \bibnamefont {Valent\'{\i}}}, \ and\ \bibinfo {author} {\bibfnamefont
  {Harald~O.}\ \bibnamefont {Jeschke}},\ }\bibfield  {title} {\enquote
  {\bibinfo {title} {{LDA$+$DMFT study of the effects of correlation in
  LiFeAs}},}\ }\href {\doibase 10.1103/PhysRevB.85.094505} {\bibfield
  {journal} {\bibinfo  {journal} {Phys. Rev. B}\ }\textbf {\bibinfo {volume}
  {85}},\ \bibinfo {pages} {094505} (\bibinfo {year} {2012})}\BibitemShut
  {NoStop}%
\bibitem [{\citenamefont {Backes}\ \emph {et~al.}(2015)\citenamefont {Backes},
  \citenamefont {Jeschke},\ and\ \citenamefont {Valent\'{\i}}}]{Backes:2015}%
  \BibitemOpen
  \bibfield  {author} {\bibinfo {author} {\bibfnamefont {Steffen}\ \bibnamefont
  {Backes}}, \bibinfo {author} {\bibfnamefont {Harald~O.}\ \bibnamefont
  {Jeschke}}, \ and\ \bibinfo {author} {\bibfnamefont {Roser}\ \bibnamefont
  {Valent\'{\i}}},\ }\bibfield  {title} {\enquote {\bibinfo {title}
  {{Microscopic nature of correlations in multiorbital
  $A{\text{Fe}}_{2}{\text{As}}_{2}$ $(A=\text{K},\text{Rb},\text{Cs})$: Hund's
  coupling versus Coulomb repulsion}},}\ }\href {\doibase
  10.1103/PhysRevB.92.195128} {\bibfield  {journal} {\bibinfo  {journal} {Phys.
  Rev. B}\ }\textbf {\bibinfo {volume} {92}},\ \bibinfo {pages} {195128}
  (\bibinfo {year} {2015})}\BibitemShut {NoStop}%
\end{thebibliography}

%

\end{document}